\documentclass[fleqn]{aa}
\pdfoutput=1

\usepackage[varg]{txfonts}

\makeatletter
\renewcommand*\aa@pageof{, page \thepage{} of \pageref*{LastPage}}
\makeatother

\usepackage{natbib,twoopt}
\usepackage[pdfborderstyle={/S/U/W 0}, linktoc=page, bookmarks=true, breaklinks=true, colorlinks=true, linkcolor=blue, urlcolor=blue, citecolor=blue]{hyperref} 
\bibpunct{(}{)}{;}{a}{}{,} 
\makeatletter
\newcommandtwoopt{\citeads}[3][][]{\href{http://adsabs.harvard.edu/abs/#3}%
{\def\hyper@linkstart##1##2{}%
\let\hyper@linkend\@empty\citealp[#1][#2]{#3}}}
\newcommandtwoopt{\citepads}[3][][]{\href{http://adsabs.harvard.edu/abs/#3}%
{\def\hyper@linkstart##1##2{}%
\let\hyper@linkend\@empty\citep[#1][#2]{#3}}}
\newcommandtwoopt{\citetads}[3][][]{\href{http://adsabs.harvard.edu/abs/#3}%
{\def\hyper@linkstart##1##2{}%
\let\hyper@linkend\@empty\citet[#1][#2]{#3}}}
\newcommandtwoopt{\citeyearads}[3][][]%
{\href{http://adsabs.harvard.edu/abs/#3}
{\def\hyper@linkstart##1##2{}%
\let\hyper@linkend\@empty\citeyear[#1][#2]{#3}}}
\makeatother

\usepackage{graphicx}	
\usepackage{amsmath, bm}	
\usepackage{amssymb}	
\usepackage{url}
\usepackage[separate-uncertainty]{siunitx}
\usepackage{multicol,multirow, makecell}
\usepackage{enumitem}
\setlist{  
  listparindent=\parindent,
  parsep=0pt,
}
\usepackage[normalem]{ulem}
\DeclareSIUnit \h {\ensuremath{h}} 

\newcommand{\thewizz}{{\tt the-wizz}}
\newcommand{\yaw}{{\tt yet\_another\_wizz}}
\newcommand{\stomp}{{\tt STOMP}}
\newcommand{\gaap}{{\sc GAaP}}
    
\begin{document}

\title{Testing KiDS cross-correlation redshifts with simulations}

\author{
  J. L. van den Busch\inst{\ref{RUB},\ref{AIfA}}
  \and
  H. Hildebrandt\inst{\ref{RUB}}
  \and
  A. H. Wright\inst{\ref{RUB}}
  \and
  C. B. Morrison\inst{\ref{Uwash}}
  \and
  C. Blake\inst{\ref{swin}}
  \and
  B. Joachimi\inst{\ref{UCL}}
  \and
  T. Erben\inst{\ref{AIfA}}
  \and\\
  C. Heymans\inst{\ref{ROE},\ref{RUB}}
  \and
  K. Kuijken\inst{\ref{leiden}}
  \and
  E. N. Taylor\inst{\ref{swin}}
}

\institute{
  Ruhr-University Bochum, Astronomical Institute, German Centre for Cosmological Lensing, Universitätsstr. 150, 44801, Bochum, Germany, \email{jlvdb@astro.rub.de}\label{RUB}
  \and
  Argelander-Institut für Astronomie, Universität Bonn, Auf dem Hügel 71, 53121 Bonn, Germany\label{AIfA}
  \and
  Department of Astronomy, University of Washington, Box 351580, Seattle, WA 98195, USA\label{Uwash}
  \and
  Centre for Astrophysics \& Supercomputing, Swinburne University of Technology, P.O. Box 218, Hawthorn, VIC 3122, Australia\label{swin}
  \and
  Department of Physics and Astronomy, University College London, Gower Street, London WC1E 6BT, UK\label{UCL}
  \and
  Institute for Astronomy, University of Edinburgh, Royal Observatory, Blackford Hill, Edinburgh, EH9 3HJ, UK\label{ROE}
  \and
  Leiden Observatory, Leiden University, P.O.Box 9513, 2300RA Leiden, The Netherlands\label{leiden}
}

\date{Received 3 July 2020 / Accepted 17 August 2020}

\abstract{
  Measuring cosmic shear in wide-field imaging surveys requires accurate knowledge of the redshift distribution of all sources.
  The clustering-redshift technique exploits the angular
  cross-correlation of a target galaxy sample with unknown redshifts
  and a reference sample with known redshifts. It represents an
  attractive alternative to colour-based methods of redshift
  calibration. Here we test the performance of such clustering redshift
  measurements using mock catalogues that resemble the Kilo-Degree Survey (KiDS). These mocks are created from the MICE simulation and closely mimic the
  properties of the KiDS source sample and the overlapping
  spectroscopic reference samples. We quantify the performance of the
  clustering redshifts by comparing the cross-correlation results with the true redshift
  distributions in each of the five KiDS photometric redshift bins. Such a comparison to an
  informative model is necessary due to the incompleteness of the reference samples at high
  redshifts. Clustering mean redshifts are unbiased at
  $|\Delta z|<0.006$ under these conditions. The redshift evolution of the galaxy bias of the
  reference and target samples represents one of the most important systematic errors when
  estimating clustering redshifts. It can be reliably mitigated at this level of precision
  using auto-correlation measurements and self-consistency relations, and will not become a
  dominant source of systematic error until the arrival of Stage-IV cosmic shear surveys. Using
  redshift distributions from a direct colour-based estimate instead of the true redshift
  distributions as a model for comparison with the clustering redshifts increases the biases in
  the mean to up to $|\Delta z|\sim0.04$. This indicates that the interpretation of clustering
  redshifts in real-world applications will require more sophisticated (parameterised) models of
  the redshift distribution in the future. If such better models are available, the clustering-redshift
  technique promises to be a highly complementary alternative to other methods of redshift calibration.
}


\maketitle



\section{Introduction}

Weak gravitational lensing (WL) experiments have been established as one of the most sensitive cosmological probes \citep[e.g.][]{Bartelmann01,Troxel18,Hikage19,Hildebrandt20a}. The aim of these experiments is to statistically probe the distribution and evolution of large-scale matter structures by studying the effect of their gravitational field on the propagation of light. The main observable effect - coherent distortions in the images of background galaxies - is very weak and can only be observed statistically from large samples of galaxies with well-measured shapes. In this regime, tight control of systematic errors is essential throughout the analysis \citep{Mandelbaum18}.

These efforts include a precise, unbiased calibration of the source redshift distribution for very large galaxy samples \citep[e.g.][]{Hoyle18,Tanaka18,Wright20a}. Complete spectroscopy is unfeasible for such surveys consisting of tens of millions of faint galaxies so that secondary estimates for the redshift distributions are required, for example redshifts from multicolour photometry known as photometric redshifts (photo-$z$; see \citealt{Salvato19} for a review). Weak lensing by the large-scale-structure of the Universe (a.k.a. cosmic shear) is a statistical effect, integrated along the line-of-sight so that redshift precision for individual galaxies is not critically important. Typically, the sources used for the measurement are divided into several so-called tomographic redshift bins. It is the redshift distributions of the sources in these bins that are required to model the observed signals. These distributions, and most importantly their mean redshifts, need to be estimated with very high accuracy, which is challenging with photometric redshifts alone due to degeneracies in colour-redshift space and incompleteness of spectroscopic template libraries.

In order to meet the stringent requirements on the accuracy of the mean redshifts
(as opposed to accuracy of the individual redshifts per se),
modern cosmic shear surveys have developed different methods of redshift calibration. If a survey contains a sub-sample of galaxies with spectroscopic redshift (spec-$z$) measurements, this sub-sample can be used to estimate the redshift distribution of the full survey. However, it is in principle required that this sub-sample is representative of the full WL source sample. If that is not the case, some re-weighting of the spec-$z$ sample can - under certain conditions - still yield an unbiased estimate of the true redshift distribution \citep{Lima08}. This re-weighting method, implemented via $k$-nearest-neighbour matching ($k$NN) or self-organised-maps (SOM), has been used widely in the WL literature \citep{Bonnett16, Masters16, Hildebrandt17, Hildebrandt20a, Wright20a, Buchs19}.

A complementary approach to estimate redshift distributions with the help of a spec-$z$ calibration sample uses galaxy angular cross-correlation measurements \citep{Newman08,Matthews10,Schmidt13,Menard13,McQuinn13}. Galaxy samples that overlap in redshift show some correlation of their angular positions on the sky. Hence, a measurement of this angular cross-correlation function can yield an estimate of the redshift distribution of an unknown sample if the other sample has accurately known redshifts. The great appeal of this method is that it does not require a reference spec-$z$ sample that is representative of the unknown target sample so long as both overlap in redshift \citep{Newman08}. For example, a bright reference sample can be used to estimate the redshift distribution of a faint target sample via cross-correlations as bright galaxies cluster with faint galaxies; they are both tracers of the same underlying matter field. This makes such a cross-correlation approach especially attractive for faint target samples that are hard to study spectroscopically in a representative way. Here we will concentrate on this approach, a method that is also dubbed `clustering redshifts'.

Any measurement of galaxy clustering - and hence also clustering redshift measurements - is affected by the unknown bias of the galaxies with respect to the underlying matter density field, which depends on galaxy type and evolves with redshift. Clustering redshifts need to be corrected for this bias before they can be used to estimate redshift distributions for cosmological inference. While the absolute value of the galaxy bias is not important for the clustering-redshift method, any redshift evolution of the galaxy bias within either the reference or the target samples will introduce a skew in the estimates of the redshift distributions. Thus, this bias evolution needs to be estimated and corrected for. While this is straightforwardly done for the reference sample, by estimating its angular auto-correlation function and exploiting its precise redshift information, such a correction is not possible for the target sample. Instead, self-consistency relations can be formulated that estimate the galaxy bias evolution of the target sample by dividing this sample into redshift slices of different width.

The aim of this work is to assess the performance of the clustering-redshift methodology and the bias corrections described above.
We use the optical and infra-red weak lensing surveys KiDS \citep[Kilo-Degree Survey,][]{Kuijken15} and VIKING \citep[VISTA Kilo Degree,][]{Edge13} as reference to create a simulated, realistic target galaxy sample, as well as a set of simulated spectroscopic calibration samples which are used as a reference in the clustering redshift measurements. These mock catalogues, based on the MICE simulation \citep[][]{Fosalba15a,Fosalba15b,Crocce15,Carretero15,Hoffmann15}, include effects such as gravitational lensing, evolving galaxy bias and spectroscopic selection effects which allows us to quantify the impact of these systematics on the clustering redshift measurements. The cross-correlation measurements are analysed with an updated methodology similar to the one presented in \citet{Hildebrandt20a}. We quantify the residual biases in such a clustering redshift experiment by comparing to the known redshift distributions of the simulated mock catalogues. This yields realistic best-practice solutions that can be used with contemporary and future cosmic shear surveys.

The paper is organised as follows. In Sect.~\ref{sec:data}, we present the simulated datasets that form the basis of this work, in particular the detailed creation of the catalogues. We note that these mock catalogues have already been used to validate the DIR and SOM redshift calibration methods in \citet{Wright20a} and \citet{Joudaki20}. The theory behind and implementation of the clustering-redshift technique for this work and differences to previous KiDS clustering redshift measurements are covered in Sect.~\ref{sec:methods}. Results are presented in Sect.~\ref{sec:results} and discussed in Sect.~\ref{sec:discussion} before the paper is summarised in Sect.~\ref{sec:summary}.


\section{Simulated data} \label{sec:data}

To assess the performance of the KiDS clustering-redshift methodology, we must construct
a simulated dataset that is similarly complex as the observational data, particularly with respect to photometric
properties and selection effects. In this section we describe the construction of such realistic mock galaxy
samples that closely resemble the KiDS+VIKING-450 \citep[KV450,][]{Wright19} cosmic shear sample
(Sect.~\ref{sec:mock_phot}) and spectroscopic reference samples (Sect.~\ref{sec:mock_spec}), which we
subsequently use to verify our clustering-$z$ methodology. We start from an existing galaxy mock catalogue
and add properties like a photometry realisation, galaxy weights and photometric redshifts (photo-$z$) in a
post-processing pipeline. This pipeline represents a blueprint for the construction of future KiDS mock samples and
is publicly available at \url{https://github.com/KiDS-WL/MICE2_mocks}.

The basis for our mock creation is the MICE simulation \citep{Fosalba15a}. MICE is a dark matter-only simulation
generated in a box of width $L = \SI{3.1}{\h^{-1} Gpc}$, thereby allowing
construction of a light cone that covers an octant of the sky. The
simulation assumes a flat $\Lambda$-CDM cosmological model, with $\Omega_{\rm m} = 0.25$, $\Omega_\Lambda = 0.75$,
$\Omega_{\rm b} = 0.044$, $\sigma_8 = 0.8$, and $h = 0.7$. The simulation traces the evolution of $\sim\SI{6.9e10}{}$ particles with mass
\SI{2.9e10}{\h^{-1} M_\odot}, from an initial redshift of $z_{\rm init} = 100$ to the present day. This high particle
density allows the simulation to match \citep[to within a few percent, ][]{Fosalba15a} theoretical predictions for 
matter clustering even on small scales ($k \sim 1\,h~\si{Mpc^{-1}}$); scales which are of particular interest for 
clustering redshift measurements. 

An additional strength of the MICE simulation is the availability of a synthetic galaxy and halo
catalogue\footnote{Distributed on \url{https://cosmohub.pic.es/} \citep{Carretero17}} for the full light-cone.
This catalogue was generated by identifying halos using a Friends-of-Friends algorithm \citep{Crocce15}, and
subsequently populating these halos with galaxies using a mixture of halo occupation distribution (HOD) and halo
abundance matching (HAM) techniques \citep{Carretero15} up to a redshift of $z \approx 1.4$.
This redshift limit implies that we are not be able to model the tails of the redshift distribution of the KV450 cosmic shear sample, which extends beyond $z = 1.4$.
However, the incompleteness of our spectroscopic reference samples (see Sect.~\ref{sec:mock_spec}) would make a clustering redshift calibration of these tails difficult.
For all analyses in this work, we use the
second version of this galaxy catalogue, which we simply refer to as MICE2. This catalogue provides galaxy positions,
shapes, stellar masses, and simulated photometry for many photometric bandpasses, such as those utilised by Euclid
\citep{Laureijs11}, Sloan Digital Sky Survey \citep[SDSS,][]{York00}, the Dark Energy Survey \citep[DES,][]{Flaugher15}, and the VIKING survey \citep{Edge13}.
As a result, the MICE2 galaxy catalogue comes pre-packaged with simulated photometry in filters similar (or identical)
to those used in KV450 ($ugriZYJHK_{\rm s}$).

As the MICE2 catalogue was constructed with gravitational lensing applications in mind, it includes both shear
(split in two components, $\gamma_1$ and $\gamma_2$) and convergence ($\kappa$) information at the position of
each galaxy \citep{Fosalba15b}. Besides the shape, gravitational lensing also affects the position
and magnifies the flux of galaxies through changes in the observed solid angle.
Therefore, both true and lensed galaxy positions are listed in MICE2. Photometry contained within the
catalogue does not, however, include a consideration for the effects of magnification. 
We utilise this fact to investigate the impact of magnification on our redshift calibration
methodology: we perform our analysis using catalogues with lensed positions and magnified fluxes, and then again
using a catalogue containing true positions and unmagnified fluxes (see Sect.~\ref{sec:systematic_results}).
Our method for implementing flux magnification is provided below.

The methods applied to the underlying dark matter-only simulation to obtain the MICE2 galaxy catalogue and our sample selection strategies (see Sect.~\ref{sec:mock_phot}) are similar to the efforts to design the Buzzard Flock synthetic sky catalogues \citep{DeRose19} for DES. Some key differences are that MICE has a higher particle density and therefore mass resolution, whereas the Buzzard mocks implement gravitational lensing with full ray-tracing opposed to MICE2 which computes lensing observables using the Born approximation.

\subsection{KV450 mock catalogue} \label{sec:mock_phot}

The combined KV450 dataset, including object detection, forced optical and infrared photometry, and photometric
redshifts (photo-$z$), is described in detail in \citet{Wright19}. The primary strength of this dataset lies with the
addition of the infrared $ZYJHK_{\rm s}$-bands, from the VIKING survey \citep{Edge13,Venemans15}, to the KiDS optical
$ugri$ dataset \citep{deJong17}, thereby significantly improving the performance of the photo-$z$ (particularly at
$z\gtrsim0.9$) obtained via template-fitting with \texttt{BPZ} \citep[Bayesian Photometric Redshift,][]{Benitez00}. 

The effective survey area of the KV450 dataset is \SI{341.3}{deg^2} \citep{Wright19}. This sample is limited to
sources with successful photometric estimates, made using the Gaussian Aperture and PSF
\citep[\gaap{},][]{Kuijken08} photometric pipeline, in all nine bands.
\gaap{} is a technique that allows to accurately measure colours by accounting for differences in the point spread function (PSF) in each
filter. This is achieved by measuring fluxes with a filter-dependent, spatially varying kernel that Gaussianises the PSF.
All sources are assigned lensing weights obtained from {\it lensfit}
\citep{Miller07,Miller13,FenechConti17,Kannawadi19}. This weighting effectively selects extended sources
with $r$-band apparent magnitudes in the interval $20 \lesssim r \lesssim 25$ \citep{Wright19}, resulting in an effective surface density \citep{Heymans12} of $n_{\rm eff} = \SI{7.38}{arcmin^{-2}}$.
Furthermore, in \citet{Hildebrandt20a} we selected sources for cosmic shear tomography as being those within the photo-$z$ window 
$0.1 < Z_{\rm B} \leq 1.2$, where $Z_{\rm B}$ is the photo-$z$ point-estimate returned by {\tt BPZ}. We split these 
sources into five non-overlapping tomographic bins with boundaries
$Z_{\rm B}\in \left\{0.1,0.3,0.5,0.7,0.9,1.2\right\}$.

The mock sample that we create based on MICE2 mimics the same selection function, object weights, and photometric
redshifts, thereby allowing us to apply the same tomographic photo-$z$ selection
on the mocks as in the real KV450 dataset. This requires us to generate photometric realisations, which match the KiDS
photometric noise properties, based on the MICE2 model magnitudes. We can do this using, per filter, the
observed median photometric depth and PSF (see Table~\ref{tab:obs_stats}).

We require only a relatively small fraction of the full MICE2 octant to match the area of the KV450 footprint. 
Due to the way in which the lightcone was constructed from the simulation box, the completeness of the MICE2 galaxy catalogue varies with position \citep[see ][]{Fosalba15a}. Therefore we select a
rectangular region between $\SI{35}{\degree} < {\rm R.A.} < \SI{55}{\degree}$ and $\SI{6}{\degree} < {\rm DEC} <
\SI{24}{\degree}$, which has a reportedly high completeness down to $i_{\rm DES} = \SI{24.0}{mag}$.
Using this subset as a basis, we construct our mock KV450 photometric sample by applying the following series of steps: application of
evolution corrections\footnote{The evolutionary correction ensures a better match in the galaxy number density between MICE2 and observational data.}, application of flux magnification, construction of photometric apertures, realisations of
photometric noise, assignment of shear-measurement weights, and computation of photometric redshifts. 
The MICE2 completeness limit, quoted above, is nominally brighter than the corresponding KV450 $i$-band magnitude limit (see Table~\ref{tab:obs_stats}).
However, the shear-measurement weights preferentially select bright objects such that the completeness limit is not an issue for our sample selection.

We start our mock sample construction by first selecting, from MICE2, the raw simulated photometry (which is noiseless,
uncorrected for evolution and magnification, and expressed in AB apparent magnitudes per filter $X$: ${\rm m}^{\rm raw}_X$) 
which uses photometric band-passes $X$ that are most similar to those used in KV450. For the OmegaCAM
$ugri$-bands and VISTA $Z$-band we use the provided SDSS $u' g' r' i' z'$-band fluxes. For the
VISTA $Y$-band we use the DES $y$-band. The VISTA $JHK_{\rm s}$-bands are provided natively within the catalogue.

Following the recommendation of \citet{Fosalba15b}, we apply a redshift dependent evolution correction to all fluxes
within the MICE2 catalogue:
\begin{equation}
  {\rm m}^{\rm evo}_{X}(z_{\rm true}) = {\rm m}^{\rm raw}_{X} - 0.8 \left[ \arctan{(1.5 \, z_{\rm true})} - 0.1489 \right],
\end{equation}
where ${\rm m}^{\rm evo}_{X}$ is the evolution-corrected apparent magnitude in filter $\rm X$, ${\rm m}^{\rm raw}_{X}$ is the raw
simulated apparent magnitude in filter $\rm X$, and $z_{\rm true}$ is the true redshift of the source. 
We then approximate the effect of magnification on our fluxes, again following the recommendation of \citet[][see
Eq.~21]{Fosalba15b}, with the correction: 
\begin{equation}
  {\rm m}^{\rm mag}_{X} = {\rm m}^{\rm evo}_{X} - 2.5 \log_{10}(1 + \delta \mu),
\end{equation}
where $\delta \mu \approx 2 \kappa$ in the weak lensing limit. We then define the `true' magnitude of each source in band
$X$, ${\rm m}^{\rm true}_{X}$, as either ${\rm m}^{\rm mag}_{X}$ (for the magnified case) or ${\rm m}^{\rm evo}_{X}$
(for the unmagnified case). 

\begin{table}[t]
  \centering
  \caption{
    Median limiting magnitudes and PSF FWHMs for each filter in the KV450 dataset.}
  \label{tab:obs_stats}
  \begin{tabular}{lcc}
    \hline\hline
    Filter & \parbox[c][][c]{20mm}{\centering PSF FWHM\\(arcsec)} & \parbox[c][][c]{30mm}{\centering \gaap{} Magnitude \\Limit ($1\sigma$, AB)} \\
    \hline
    $u$         & 1.0 & 25.5 \\
    $g$         & 0.9 & 26.3 \\
    $r$         & 0.7 & 26.2 \\
    $i$         & 0.8 & 24.9 \\
    \hline
    $Z$         & 1.0 & 24.9 \\
    $Y$         & 1.0 & 24.1 \\
    $J$         & 0.9 & 24.2 \\
    $H$         & 1.0 & 23.3 \\
    $K_{\rm s}$ & 0.9 & 23.2 \\
    \hline
  \end{tabular}
\end{table}

\begin{figure*}[t]
  \includegraphics[width=0.97\columnwidth]{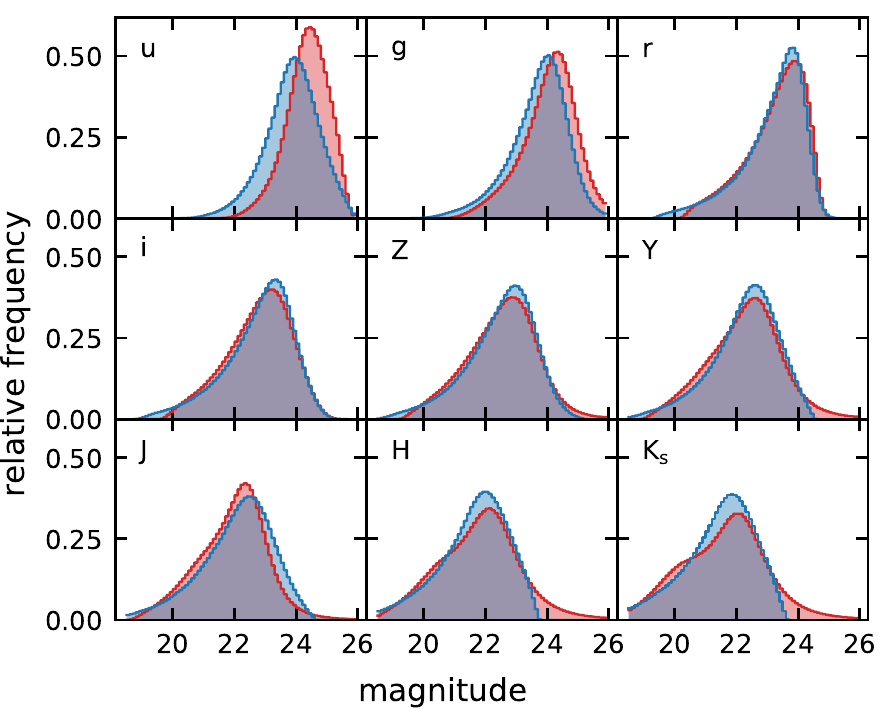}
  \hfill
  \includegraphics[width=0.97\columnwidth]{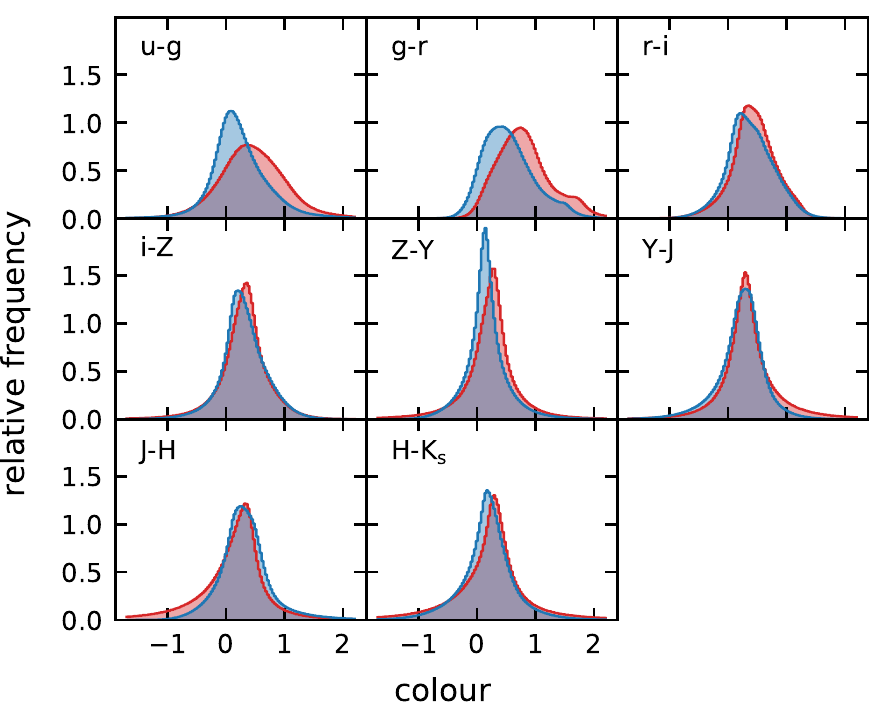}
  \caption{
    Comparison of the colours and magnitudes for the KV450 (red) and MICE2 mock (blue) cosmic shear samples. All histograms are
    unweighted, but objects with zero {\it lensfit} weight are excluded from both samples. The simulated distributions match the
    data well, except that MICE2 fluxes tend to be bluer overall.}
  \label{fig:MICE2_mags_colours}
\end{figure*}

We now derive `observed' photometric quantities (magnitudes and uncertainties) per-filter, by adding representative
photometric noise to the values of ${\rm m}^{\rm true}_X$. This requires us to match the photometric signal-to-noise (SN) of the simulations,
again per-filter, to that which is observed in KiDS and VIKING imaging.
In order to simulate realistic photometric errors, we need to estimate an effective aperture size for each galaxy. We first derive the
half-light radius $R_{{\rm E},i}$ from the tabulated disk and bulge radii as well as the bulge-to-total flux ratio of the two component 
\citet{Sersic63} profiles of the MICE2 galaxies.
Based on this projected galaxy size we compute a corresponding photometric aperture the KiDS pipeline would apply:
The per-filter aperture major ($a^{\rm ap}_{X,i}$) and minor axes ($b^{\rm ap}_{X,i}$) are
\begin{align}
  \label{eq:aperture_a}
  a^{\rm ap}_{X,i} &= \sqrt{\sigma_{{\rm PSF},X}^2 + (2.5 R_{{\rm E},i})^2}  \qquad \text{and} \\
  \label{eq:aperture_b}
  b^{\rm ap}_{X,i} &= \sqrt{\sigma_{{\rm PSF},X}^2 + \left(2.5 \frac{b^{\rm int}_i}{a^{\rm int}_i} R_{{\rm E},i}\right)^2},
\end{align}
where $\sigma_{{\rm PSF},X}$ is the filter $X$ PSF standard deviation (derived from the FWHM, see
Table~\ref{tab:obs_stats}), and the intrinsic major-to-minor axis ratio $a^{\rm int}_i/b^{\rm int}_i$ of source $i$.
Using the aperture area $A_{X,i}^{\rm ap} = \pi a^{\rm ap}_{X,i} b^{\rm ap}_{X,i}$, the SN ratio values of ${\rm SN}_{X,i}$ can now be computed as:
\begin{equation}\label{eq:SN}
  {\rm SN}_{X,i} = 10^{-0.4 \left({\rm m}^{\rm true}_{X,i} - {\rm m}^{\rm lim}_X\right)} \sqrt{\frac{\pi \sigma_{{\rm PSF},X}^2}{A^{\rm ap}_{X,i}}} \, k \,,
\end{equation}
where ${\rm m}^{\rm lim}_X$ is the observed \gaap{} magnitude limit in filter $X$ (see Table~\ref{tab:obs_stats}) and $k$ is a free scaling parameter. We compare the distributions of ${\rm SN}_{X}$ in the simulation with the observed signal-to-noise distributions, as determined from the correspondingly measured \gaap{} magnitude.
By varying the scaling parameter, we find that a good match between the simulated and observed distributions is achieved at $k \approx 1.5$ in all filters.
We then compute Gaussian uncertainties in magnitude and generate observed magnitudes:
\begin{equation}
  {\rm m}^{\rm obs}_{X,i} = {\rm m}^{\rm true}_{X,i} + x  \quad \text{with}~ x \sim \mathcal{N}\left(0,\frac{2.5}{\ln{10}}\frac{1}{{\rm SN}_{X,i}}\right) \,.
\end{equation}
We use these magnitudes in Eq.~\ref{eq:SN}, substituting ${\rm m}^{\rm true}_{X,i}$ with ${\rm m}^{\rm obs}_{X,i}$, to
derive the observed SN per filter and source, ${\rm SN}^{\rm obs}_{X,i}$. Finally, we perform a quasi-detection in each
band, only keeping magnitude information for sources with ${\rm SN}^{\rm obs}_{X,i} > 1.0$.
We note that the precise implementation of this detection limit is not critical due to the implementation of the shape measurement weights that limit our analysis to significantly higher $\rm SN$ anyway.

All cosmic-shear sources in KV450 have an associated shape-measurement weight, which is assigned based on the
confidence of their shape measurement. As we do not have simulated imaging for the MICE2 dataset, we cannot recreate
this value from the simulations directly. Instead we exploit the fact that these weights are strongly correlated with
magnitude (particularly the $r$-band magnitude, as this is the band which is used for galaxy shape measurement).
Therefore, to assign shear measurement weights to simulated sources, we simply perform a $k$NN matching in 9-dimensional magnitude space
between the simulated and real photometric datasets; the simulated data then inherit the shear weight of
the nearest neighbour in the data. In cases where there is no nearest neighbour match within a \SI{1.0}{mag} Euclidean 
radius, we assign a weight of zero\footnote{This situation arises in about \SI{2}{\percent} of the cases, in particular for the faintest galaxies with $r \gtrsim 24$.}.
This assignment produces well behaved shape-measurement weights, while also encoding
any strong selection effects that are present in the data but overlooked in our simulation setup.

The resulting colour and magnitude distributions for the mock and real photometric data are shown in
Fig.~\ref{fig:MICE2_mags_colours}. These distributions match the data fairly well, except for the $u$ and
$g$-bands; MICE2 galaxies are typically bluer than we see in the data. The mock galaxy magnitudes do not reproduce the
faint tails of the near-infrared photometry seen in KV450, which we attribute to stronger depth variations
seen in the VIKING data, which are not modelled in our analytic prescription (Eq.~\ref{eq:SN}). 

\begin{figure}[h]
  \includegraphics[width=0.95\columnwidth]{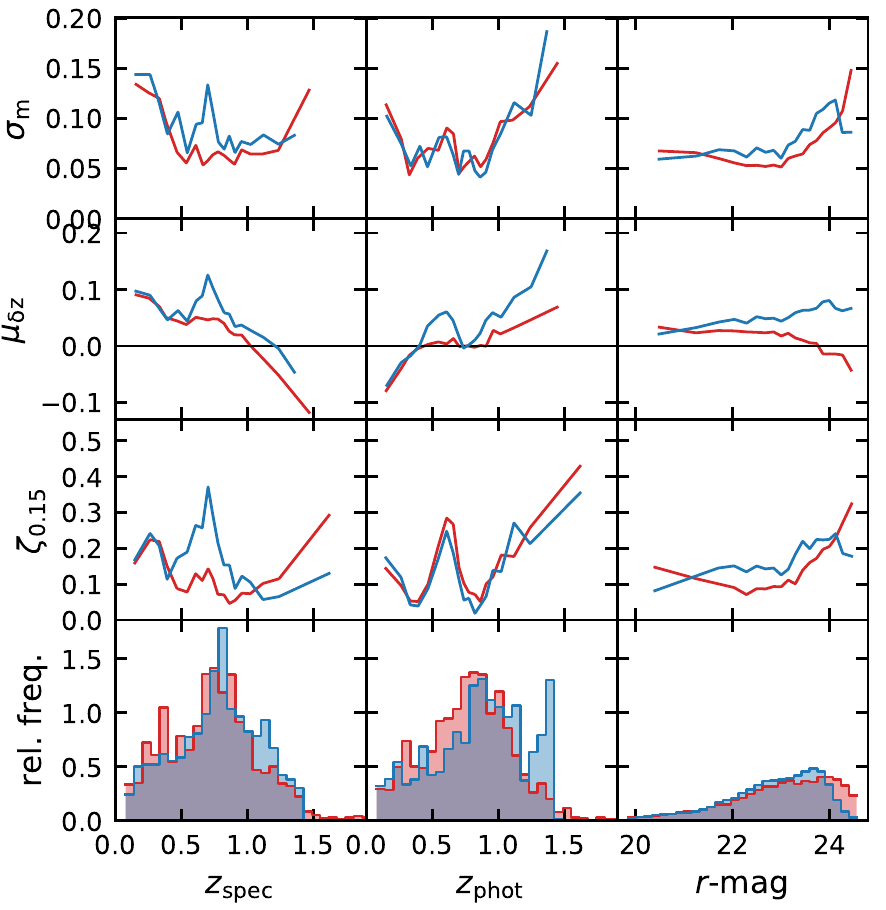}
  \caption{
    Comparison of statistics of the scaled photo-$z$ bias $\delta z = (z_{\rm B} - z_{\rm spec}) / (1 + z_{\rm spec})$
    of the KV450 (red) and MICE2 mock (blue) datasets. The galaxies used for this compilation originate
    from DEEP2, VVDS, and zCOSMOS. The columns from left to right show statistics as a function of spectroscopic
    redshift, photometric, and $r$-band magnitude (shown as histograms in the bottom row). The rows from top to bottom
    show the median-absolute-deviation ($\sigma_{\rm m}$), mean ($\mu_{\delta z}$), and the outlier fraction
    ($\zeta_{0.15}$, the fraction of objects with $\delta z > 0.15$).}
  \label{fig:MICE2_zstat}
\end{figure}

Finally, we perform a photo-$z$ estimation for the simulated data using our observed magnitudes and magnitude
uncertainties. To ensure consistency with the data, we implement the same photo-$z$ code (i.e. {\tt BPZ}) and setup; 
that is, we use the same redshift prior \citep{Raichoor14} and template set \citep{Capak04} as described in
\citet{Wright19}. A difference between the simulations and the data, however, lies in the redshift priors applied to the
simulations. As we know {\it a priori} that the mock catalogue only contains galaxies within $0.07 \lesssim z\lesssim 1.41$,
we limit the \texttt{BPZ} redshift-prior to this range. We also reject objects if their aperture
photometry cannot be measured in the $i$-band, the reference magnitude for the redshift prior, to avoid spurious
photo-$z$ estimates.

The resulting mock photometric redshift properties are well matched to the KV450 data as shown in 
Fig.~\ref{fig:MICE2_zstat}. Following \citet{Wright19}, we measure statistics of the scaled photo-$z$ bias distribution, 
$\delta z = (z_{\rm B} - z_{\rm spec}) / (1 + z_{\rm spec})$, as a function of spectroscopic redshift, photometric redshift, 
and $r$-band apparent magnitude. We find agreement between the mocks and data in each of the test statistics explored: 
distribution scatter ($\sigma_m$; estimated using the normalised median-absolute-deviation from median, nMAD), mean bias
($\mu_{\delta z}$), catastrophic outlier fraction ($\zeta_{0.15}$; the fraction of sources with $|\delta z|>0.15$), and
relative counts. 
In particular, we highlight the similarities between the scatter and outlier fractions seen as a function of photo-$z$,
which are closely matched between the mocks and the data.

\subsection{Spectroscopic mock catalogues} \label{sec:mock_spec}

The KV450 photometric dataset has spatial overlap with a rich set of spectroscopic surveys, which are needed
for optimal redshift calibration. For a detailed discussion of all available spectroscopic data which intersect
the KV450 footprint, we refer the interested reader to \citet{Hildebrandt20a}. Here we briefly summarise the
spectroscopic selection functions and how to implement these on the MICE2 mocks for the subset of the datasets
which we use for clustering redshift measurements. These samples can be divided in two categories.
First, there are four wide-area spectroscopic surveys with a combined overlap with KV450 of \SI{212.0}{deg^2}, namely
the Sloan Digital Sky Survey \citep[SDSS,][]{Alam15},
the 2-degree Field Lensing Survey \citep[2dFLenS,][]{Blake16},
the Galaxy and Mass Assembly \citep[GAMA,][]{Driver11} survey,
and the WiggleZ Dark Energy Survey \citep[WiggleZ,][]{Drinkwater10}.
Secondly, KiDS relies on deep spectroscopic surveys, which each have a post-masking overlap area with KiDS of less than
\SI{1}{deg^2}, to calibrate the high-redshift portions of the redshift distributions;
the DEEP2 Galaxy Redshift Survey \citep[DEEP2,][]{Newman13},
the VIMOS VLT Deep Survey \citep[VVDS,][]{LeFevre13},
and the Cosmic Evolution redshift survey \citep[zCOSMOS,][]{Lilly09}.
As with the KV450 mock galaxy sample, we aim to create mock spectroscopic samples with similar complexity as seen in 
the actual spectroscopic data.

\begin{figure}[t]
  \centering
  \includegraphics{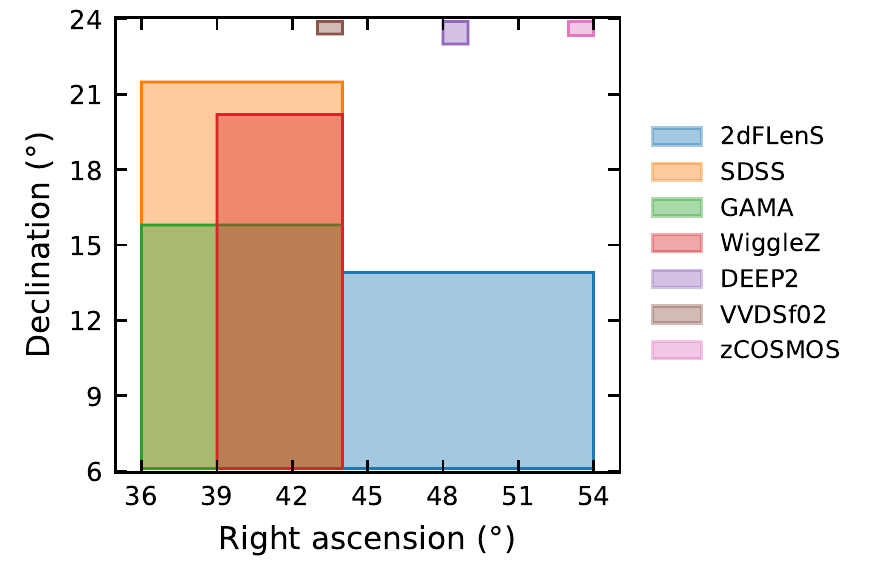}
  \caption{
    The distribution of the mock spectroscopic samples within the mock KV450 footprint (indicated by the axes limit).}
  \label{fig:zspec_footprints}
\end{figure}

We start our mock spectroscopic source generation by simplifying the complex footprints of each of our spectroscopic
samples by selecting rectangular regions which respect their size and the various overlaps between each other and with the true KV450 data.
These footprints can be seen in Fig.~\ref{fig:zspec_footprints}, demonstrating the various spatial overlaps
between the different samples in our spectroscopic compilation and matching (in area) those of the corresponding
data samples within a few percent.

Next, we apply the spectroscopic target selection functions (where possible) and adjustments (where
necessary) to obtain the closest match possible between our mock and data spectroscopic redshift distributions. 
These selection functions are all magnitude- and/or colour-dependent, and are variously based on imaging that is either 
shallower (2dFLenS, GAMA, SDSS, and WiggleZ) or deeper (DEEP2, VVDS, and zCOSMOS) than the KV450 imaging. As a result,
we are able to use the mock KV450 photometry (i.e. ${\rm m}^{\rm obs}_{X,i}$) to perform spectroscopic selections for
the former samples, but revert to the noiseless photometry (i.e. ${\rm m}^{\rm mag}_{X,i}$) for the 
selection of the latter samples. A tabular summary of these selection functions and additional plots for VVDS and zCOSMOS can be found as supplementary material in Appendix~\ref{sec:mock_selection}.

\subsubsection{SDSS}

SDSS covers large parts of the northern KiDS fields. Our sample consists of
galaxies from the SDSS Main Galaxy Sample \citep{Strauss02}, BOSS \citep[Baryon Oscillation Spectroscopic
Survey,][]{Dawson13}, and the QSO sample \citep{Schneider10} at high redshifts.
The selections applied to MICE2 for each of these samples are given in Table~\ref{tab:spec_SDSS}, and
differences with respect to the literature are justified in the following.

First, we approximate the main galaxy sample by selecting
objects with ${\rm m}^{\rm obs}_{r,i} < 17.7$, which is slightly brighter than the literature limit of $17.77$ 
\citep{Strauss02}. We attribute this small discrepancy to the different magnitude definitions used in SDSS and our
mock KV450 dataset (i.e. simple Petrosian vs our $2.5R_e$ aperture fluxes). Our updated selection gives a better match 
between the mock and simulated redshift distributions for the SDSS main galaxy sample. 

\begin{figure}[t]
  \centering
  \includegraphics[width=0.93\columnwidth]{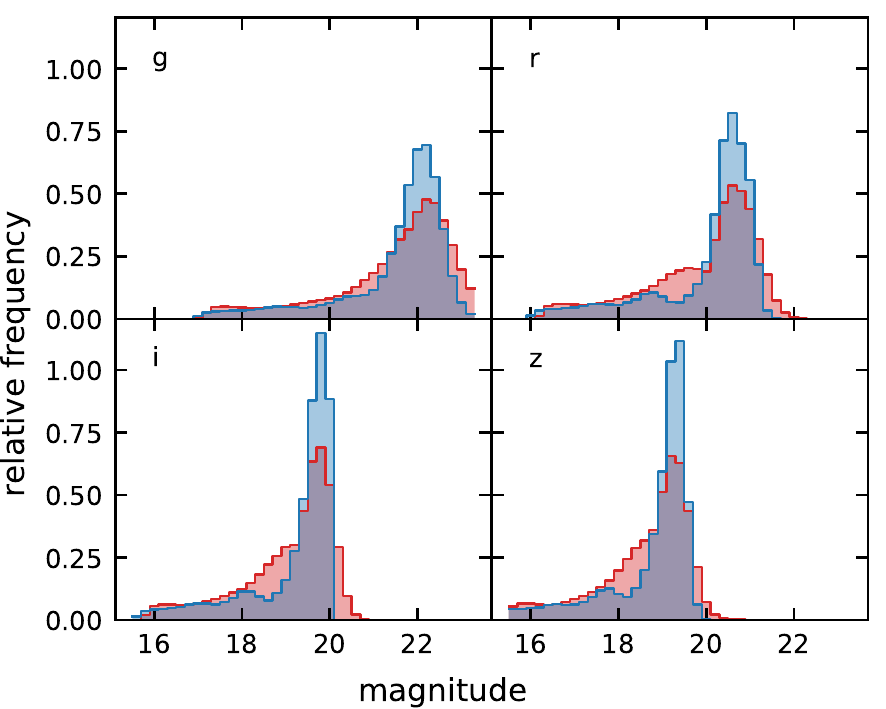}
  \\\vspace{2mm}
  \includegraphics[width=0.93\columnwidth]{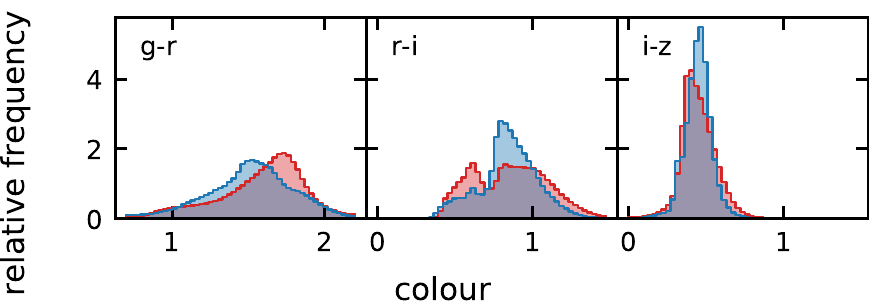}
  \\\vspace{2mm}
  \includegraphics[width=0.93\columnwidth]{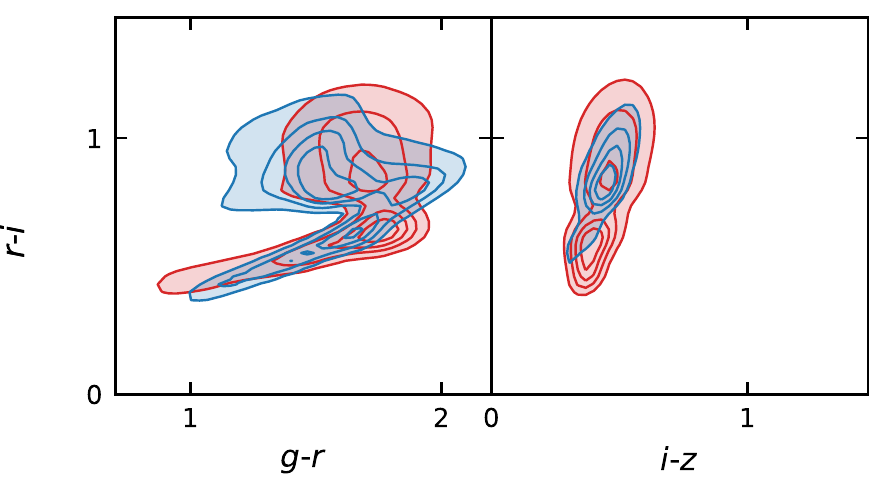}
  \caption{
    Comparison of colour and magnitude distributions for the simulated (blue) and observed (red) BOSS LOWZ and CMASS
    datasets.}
  \label{fig:BOSS_props}
\end{figure}

The BOSS LOWZ and CMASS samples use a set of auxiliary colours, based on $gri$-band model magnitudes and optimised for the
selection of luminous red galaxies (LRGs), which are defined as: 
\begin{align}
  \label{eq:BOSS_cut1}
  c_\parallel &= 0.7\, (g-r) + 1.2\, (r-i - 0.18); \\
  \label{eq:BOSS_cut2}
  c_\perp &= (r-i) - (g-r) / 4 - 0.18; {\rm and} \\
  \label{eq:BOSS_cut3}
  d_\perp &= (r-i) - (g-r) / 8. 
\end{align}
These auxiliary colours can be directly computed for our MICE2 sources given the available bandpasses:
${\rm m}^{\rm obs}_{X,i}$ for $X \in gri$.  
We are therefore able to apply the literature colour-cuts for the LOWZ and CMASS samples \citep{Dawson13}, albeit
again with some minor modifications. These modifications are applied to construct better matches between the
simulations and data in colour and magnitude space: see Fig.~\ref{fig:BOSS_props} for a comparison of these values
between our mocks and the data. 

\begin{figure}[t]
  \centering
  \includegraphics[width=0.95\columnwidth]{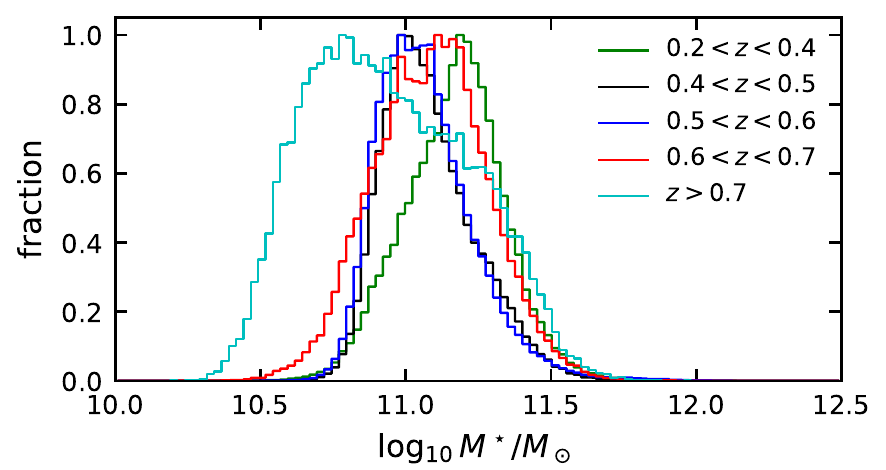}
  \caption{
    The MICE2 stellar mass distributions for the BOSS mock sample in bins of redshift, normalised to their peak value. The comparison to Fig.~10 in \citet{Maraston13} reveals significantly lower stellar masses in the highest redshift bin.}
  \label{fig:BOSS_Mstellar}
\end{figure}

To verify the fidelity of our BOSS selection, we compare the distributions of stellar mass, in bins of redshift,
between our mock sample (see Fig.~\ref{fig:BOSS_Mstellar}) and those of \citet[][see their Fig.~10]{Maraston13}. 
At redshifts below $z \sim 0.6$ our BOSS sample is in good agreement with \citet{Maraston13}, except for a constant 
\SI{0.2}{dex} systematic offset between the two samples. This systematic offset is not surprising, given the 
systematic differences between the \citet{Maraston13} stellar population synthesis models and those of
\citet{Bruzual03}, which are used to estimate MICE2 stellar masses. 
Sources with $z > 0.7$ in our mock sample, however, have a significantly lower stellar mass than the corresponding 
BOSS galaxies in \citet{Maraston13}. Whereas in
\citet{Maraston13} this redshift bin has the highest mean stellar mass, we find that high-redshift MICE2 BOSS
galaxies have the lowest mean stellar mass; an offset of $\sim0.5$ dex in mass. This in turn suggests that 
the biasing of these galaxies will differ in the simulations compared to the data. 
We expect this to have a minor impact on the results presented here. It could however be of importance in
other applications of this dataset which are more sensitive to the absolute value of the galaxy bias. 

There is no obvious way to implement the SDSS QSO sample within MICE2, as the simulation does not include active
galactic nuclei in the galaxy catalogue construction. We therefore approximate the selection of quasars using physical
properties. We assume that quasars are triggered exclusively in dense environments ($M_{\rm halo}
>\SI{1e13}{M_\odot}$), are always hosted by central galaxies (${\tt flag\_central}=1$), and their hosts have high stellar masses $M_{\rm
stellar} > \SI{1e11}{M_\odot}$. This selection yields a sample with both a number
density and redshift distribution similar to those found in the observed QSO sample. 

\subsubsection{2dFLenS}

2dFLenS is similar to the BOSS sample, but occupies the southern KiDS fields and consists of three subsamples
with different selection functions: the low-$z$, the mid-$z$, and the high-$z$ sample.
The selection criteria for each of these samples are given in Table~\ref{tab:spec_2dFLenS}. 
The low-$z$ and mid-$z$ selections are very similar to the BOSS LOWZ and CMASS selections, and are based on 
the same complement of $gri$-magnitudes and auxiliary colours 
\citep[i.e. Eq.~\ref{eq:BOSS_cut1}-\ref{eq:BOSS_cut3},][]{Blake16}. We apply these cuts again with
minor parameter fine-tuning to better reproduce the observed redshift and colour/magnitude distributions. 
The high-$z$ sample, however, contains additional selections incorporating fluxes from the Wide-field Infrared Survey
Explorer \citep[WISE,][]{Wright10} $W1$-band (central wavelength $\bar{\lambda}=3.6\mu m$), which is not available in
MICE2. We approximate this selection using the VIKING $K_{\rm s}$-band (central wavelength $\bar{\lambda}=2.15\mu m$)
in place of $W1$.
While not exact, comparisons between the redshift distributions in MICE2 and the 2dFLenS data
suggest that this approximation is suitable for the selection accuracy required in this work.
Furthermore, 2dFLenS applies a density downsampling after selecting targets according to the criteria in Table~\ref{tab:spec_2dFLenS}, resulting in approximately half the density of the corresponding BOSS LRG samples.
This downsampling is partly comprised in our modified selection, which yields an approximately $\SI{30}{\percent}$ higher density than the 2dFLenS data sample.
We consider this difference sufficiently small for our purposes that we do not implement an additional downsampling for the MICE2 sample.

\begin{figure}[t]
  \centering
  \includegraphics[width=0.93\columnwidth]{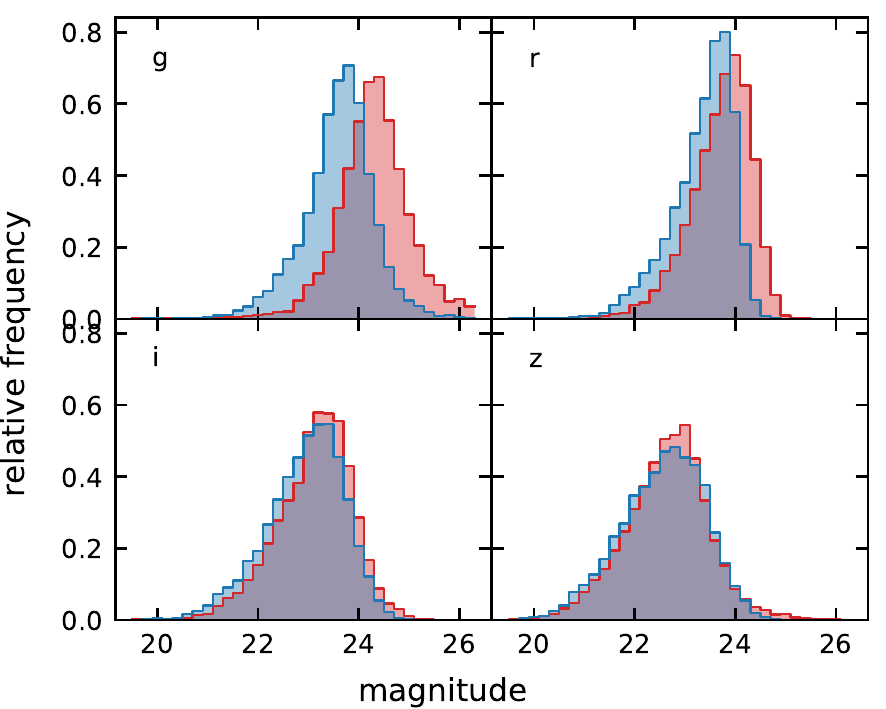}
  \\\vspace{2mm}
  \includegraphics[width=0.93\columnwidth]{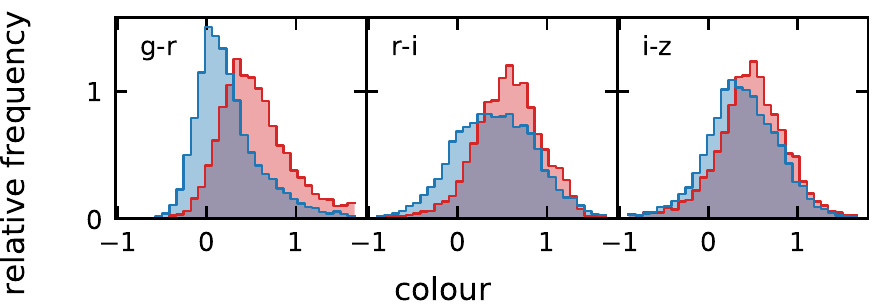}
  \\\vspace{2mm}
  \includegraphics[width=0.93\columnwidth]{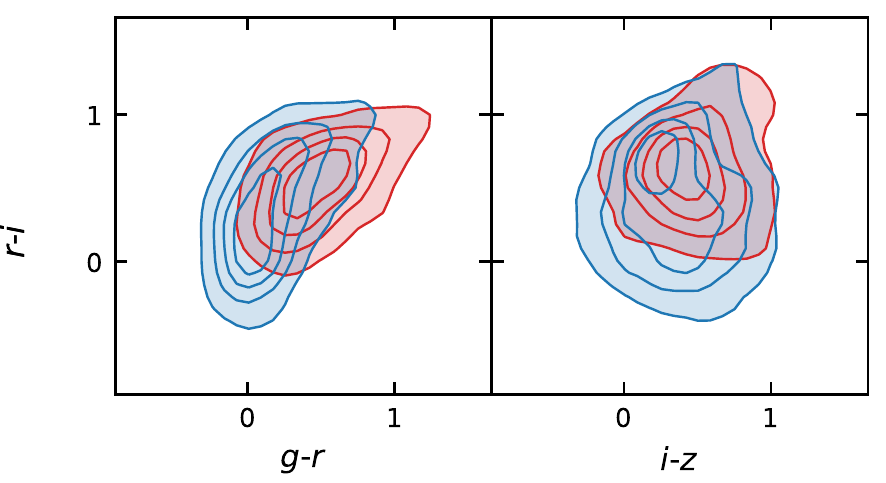}
  \caption{
    Comparison of colour and magnitude distributions for the simulated (blue) and observed (red) DEEP2 datasets.}
  \label{fig:DEEP2_comp}
\end{figure}

\subsubsection{GAMA}

GAMA is a flux-limited spectroscopic survey  distributed over three equatorial fields that
overlap with the northern KiDS fields. The survey data used here \citep{Driver11,Liske15} is highly complete 
($>\SI{98}{\percent}$) to the magnitude limit of $r = \SI{19.8}{mag}$, with the bulk of galaxies residing at redshifts of
$z_{\rm spec} \lesssim 0.4$. We find that a single $r < 19.87$ selection is best suited to reproduce the redshift distribution of GAMA in our MICE2 mocks. 

\subsubsection{WiggleZ}

WiggleZ is the deepest of the wide area surveys overlapping with KV450, extending
to $z_\mathrm{spec} \lesssim 1.1$. This dataset spans the gap between the wide and deep spectroscopic datasets.
The WiggleZ selection function \citep{Drinkwater10} consists of a number of cuts which both include and exclude
certain parts of the far-UV to near-IR colour-colour space, and are summarised in 
Table~\ref{tab:spec_WiggleZ}. As we do not have mock UV photometry in the MICE2 catalogues, we are unable to fully reproduce the literature selection function of
WiggleZ. This deficiency means that our initial
selection function produces a markedly different redshift distribution in the mocks compared to the data. To accurately
model the redshift distribution of WiggleZ we are therefore required to - after performing all the possible
selections - apply a direct matching of the simulation and data redshift distributions. We do this via a direct
redshift-dependent down-sampling of our initial WiggleZ sample to match the simulations to the data. 

\subsubsection{DEEP2}

\begin{figure}[t]
  \centering
  \includegraphics{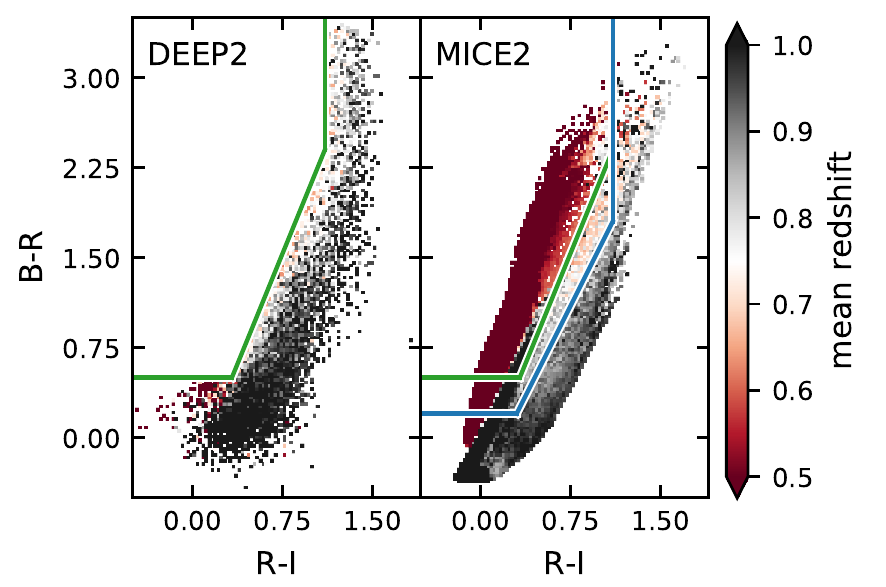}\\
  \caption{
    Left: DEEP2 spectroscopic data in $R-I$-$B-R$ colour space with the original target selection indicated in green, right:
    MICE2 galaxies (based on noiseless model magnitudes) with the original and the fiducial data selection indicated in blue.}
  \label{fig:DEEP2_selection}
\end{figure}

\begin{figure}[t]
  \centering
  \includegraphics{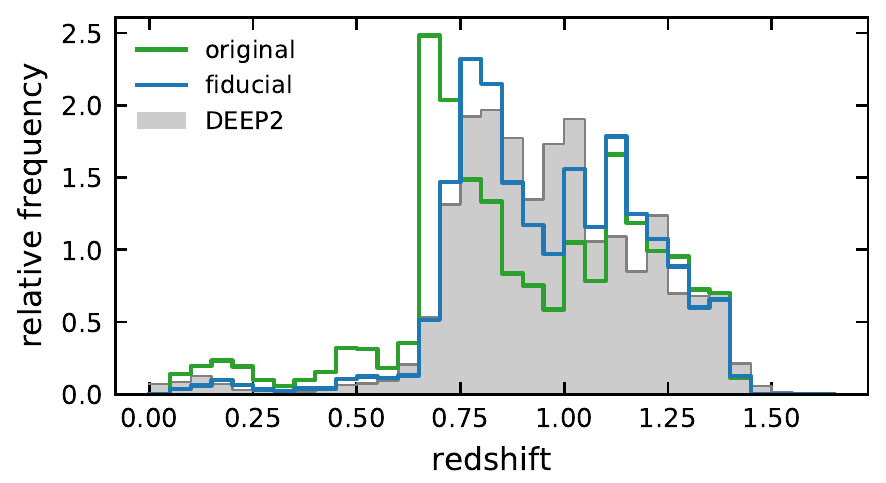}
  \caption{
    The DEEP2 redshift distribution in comparison to the redshift distribution of MICE2 galaxies selected with the
    original colour cut \citep[][green]{Newman13} and our fiducial colour cut (blue)}
  \label{fig:DEEP2_nz}
\end{figure}

DEEP2 has been covered by KiDS- and VIKING-like observations in two fields at $\mathrm{R.A.} \approx \SI{02}{h}$
and \SI{23}{h}. Sources in these fields are pre-selected by colour to target galaxies in the range
$0.75 \lesssim z_\mathrm{spec} \lesssim 1.5$. This is achieved by cuts in the Johnson $B$-$R$, $R$-$I$ colour-colour 
space \citep{Newman13}. These selections are listed in Table~\ref{tab:spec_DEEP2},
and the resulting colour- and magnitude-distributions are shown in Fig.~\ref{fig:DEEP2_comp}. Johnson magnitudes are available within MICE2, and so we are
able to apply this selection directly to the simulations. When we apply these cuts to MICE2, however, we find significant
contamination by low redshift objects in the sample, and a clear shift of the cut-off redshift from $z > 0.75$ to 
$z > 0.65$. Informed by the redshift distribution of the $B$-$R$, $R$-$I$ colour-colour space within MICE (shown
graphically in Fig.~\ref{fig:DEEP2_selection}), we adjust two of the colour cuts to obtain redshift and colour distributions 
that are closer to those observed in the DEEP2 observations. The influence of our updated colour-cuts in DEEP2 can be
seen in Fig.~\ref{fig:DEEP2_nz}, which shows clearly that our updated colour cuts yield a closer match to the observed
DEEP2 redshift distribution. We hypothesise that the need for different
colour cuts in MICE2 is partially due to our use of noiseless magnitudes in the sample definition, but moreover is
because of an extension of the colour-redshift space, at intermediate redshifts, to more negative colours than is seen
in the data. Finally, we model spectroscopic incompleteness within the DEEP2 sample using the 
redshift-completeness function presented in \citet{Newman13}, which shows a clear decrease in the fraction of sources
with high-confidence ($nQ\ge3$) redshifts with increasing $R$-band magnitude. Finally, we 
randomly downsample the remaining sample to $\sim \SI{60}{\percent}$ of its initial size, to obtain a
similar number of objects as found in the observational data. 

\subsubsection{VVDS}

The VVDS field at $\mathrm{R.A.} \approx \SI{02}{h}$ is selected via a combination of both a wide-field selection
and a deep-drill selection realised by simple magnitude cuts which add additional spectra over the whole
$0 < z_\mathrm{spec} \lesssim 1.3$ redshift range. However, the deep sample is overwhelmingly dominant in this
field, and so we opt to simulate this selection only. The deep sample is defined by a simple Johnson $I$-band
magnitude limit, $I < 24.0$. We find that implementing
this limit in the simulations, without modification, results in a sample well matched to the observations. 
We implement the literature spectroscopic success rate for VVDS \citep[][Figs.~13.a/b, 16]{LeFevre05} as a function of
both $I$-band magnitude and true-redshift. These selections are performed independently; that is, we assume no
correlation between the sources removed in the selection. Finally, the VVDS spectroscopic sample has a roughly \SI{25}{\percent}
completeness in the 2hr field; we find, however, that a $\sim\SI{17}{\percent}$ completeness is required to reproduce (in MICE2)
the number density of VVDS spectra seen in the observations. Hence, we down-sample the VVDS mock catalogue to that density. A comparison of
observed and simulated colours/magnitudes in VVDS can be found in Fig.~\ref{fig:VVDSf02_comp}.

\subsubsection{zCOSMOS}

As with the VVDS sample, our zCOSMOS sample is a combination of two distinct subsamples: the public `bright'
and a proprietary `deep' sample.
The deep sample in zCOSMOS preferentially targets objects at $z > 1.5$, which is beyond the maximum redshift 
available in MICE2. Therefore, we opt to simulate only the bright selection, which is defined by Johnson $I$-band
magnitudes in the range $15.0 < I < 22.5$. We apply this selection as-is to MICE2, finding good agreement in the
colours and redshift distributions between the simulations and the observations. We apply the spectroscopic
success rate for zCOSMOS, which is given as a joint function of $I$-band magnitude and
redshift \citep[][Fig.~3]{Lilly09}. Finally, we randomly down-sample the resulting spectra by \SI{33}{\percent}, 
to match the observed zCOSMOS spectroscopic number density. Figure~\ref{fig:zCOSMOS_comp} shows zCOSMOS and our mock
sample in colour and magnitude space. We note the absence of the deep sample creates a clear dearth of spectra at faint
magnitudes, but does not seem to systematically bias the colour-colour space in our zCOSMOS simulation. 

\subsubsection{Idealised spectroscopic sample}

\begin{figure}[t]
  \centering
  \includegraphics{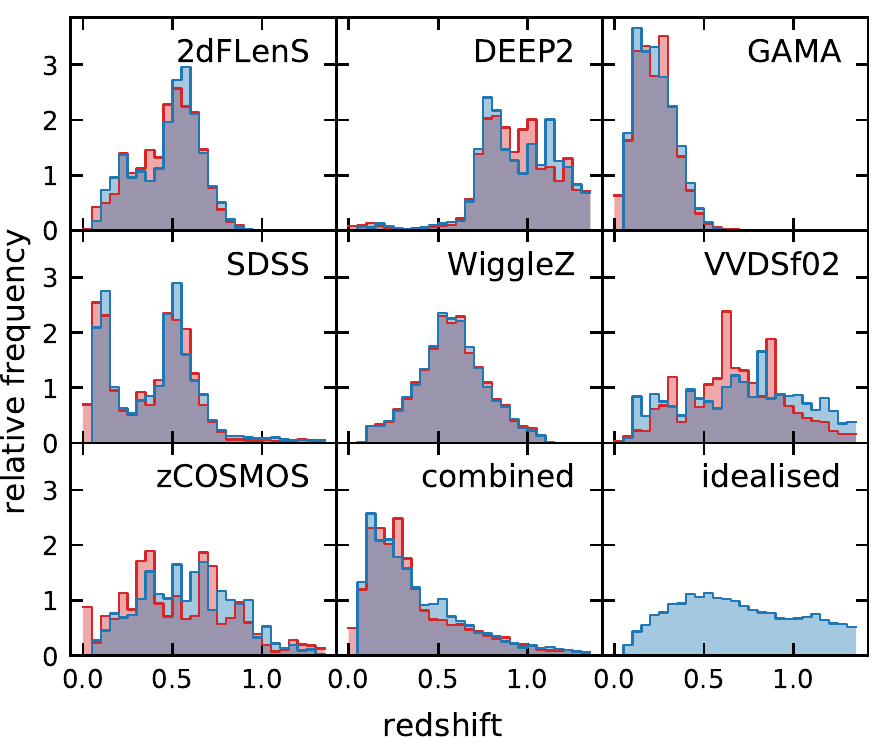}
  \caption{
    The redshift distributions of the spectroscopic data catalogues (red) compared to their corresponding MICE2 mock
    counterparts (blue). The idealised sample in the bottom right panel is used for benchmarking purposes only.}
  \label{fig:zspec_comparison}
\end{figure}

Finally we create an idealised mock spectroscopic sample, defined by selecting MICE2 sources with $r < 24.0$
that lie within the footprint of our 2dFLenS and SDSS mock galaxy samples. This sample is then sparse-sampled to a
number density that is $\sim \SI{10}{\percent}$ of the mock KV450 number density. 
Such an idealised sample allows the computation of clustering
redshifts for our mock KV450 sample excluding the influence of spectroscopic selection functions, and therefore allows
us to estimate the influence of these selections on our results. 

The final redshift distributions of all our mock spectroscopic compilations are shown in
Fig.~\ref{fig:zspec_comparison}. The mock redshift distributions agree well with their data counterparts
for both mean and median statistics (excluding those whose data distributions exhibit
considerable tails above ${\rm max}\left(z_{\rm sim}\right) = 1.4$) to within
$|\Delta z|\lesssim 0.05$ (see Table~\ref{tab:spec_stats}).
These samples are quite different from \citet{Scottez18}, who also use MICE2 to study the performance of clustering redshifts, since the ones presented here reflect some of the complications that arise in practice with spectroscopic reference samples, such as realistic spatial overlaps or redshift incompleteness, which results in more complex selection functions.

\begin{table}
  \centering
  \caption{
    Comparison of galaxy densities and median redshifts of the spectroscopic data and mock samples.}
  \label{tab:spec_stats}
  \renewcommand{\arraystretch}{1.2}
  \begin{tabular}{lrrrr}
    \hline\hline
    Survey & \makecell{$n_{\rm MICE}$ /\\ \si{arcmin^{-2}}} & \makecell{$n_{\rm spec}$ /\\ \si{arcmin^{-2}}} & $z_{\rm MICE}^{\rm median}$ & $z_{\rm spec}^{\rm median}$ \\
    \hline
    2dFLenS    & 0.020 & 0.015 & 0.51 & 0.50 \\
    GAMA       & 0.289 & 0.292 & 0.21 & 0.22 \\
    SDSS (all) & 0.052 & 0.047 & 0.46 & 0.54 \\
    WiggleZ    & 0.063 & 0.072 & 0.57 & 0.57 \\
    \hline
    DEEP2      & 2.96\hphantom{0} & 2.96\hphantom{0} & 0.96 & 0.96 \\
    VVDS-02h   & 2.54\hphantom{0} & 2.52\hphantom{0} & 0.74 & 0.69 \\
    zCOSMOS    & 4.56\hphantom{0} & 4.58\hphantom{0} & 0.59 & 0.52 \\
    \hline
  \end{tabular}
  \renewcommand{\arraystretch}{1.0}
\end{table}


\section{Redshifts from cross-correlations} \label{sec:methods}

Due to the gravitational clustering of matter in the Universe, the positions of objects which reside in a common volume (i.e. in the same large-scale structure) 
are highly correlated. Conversely, the positions of objects from disparate volumes/structures are uncorrelated \citep[except for a small
contribution from magnification, see e.g.][]{Gatti18}. We can use this fact to constrain the redshift distributions of an ensemble of extragalactic
sources by measuring the amplitude of their angular cross-correlation with a tracer sample of galaxies with known redshifts. This
approach to redshift distribution estimation is known as `cross-correlation redshifts' or simply `clustering redshifts'.

\subsection{Basic formalism} \label{sec:formalism}

In the literature there are several different approaches to clustering redshift estimation 
\citep[e.g.][]{Newman08,McQuinn13,Menard13}. In this work we follow an approach similar to \citet{Johnson17}, described
briefly here. 

Consider two samples of extragalactic objects which overlap in three dimensions:
\begin{enumerate}
  \item A reference sample ($s$) with known angular positions and redshift distribution $n_{\rm s}(z)$ obtained from
  secure point redshift estimates (typically spectroscopic redshifts).
  \item A target sample ($p$), also with known angular positions but without precise redshift information (typically selected photometrically); the 
  redshift distribution of this sample, $n_{\rm p}(z)$, is what we wish to recover.
\end{enumerate}
The angular cross-correlation of sources within the reference and target samples, at a fixed reference-sample redshift
$z$ and separation angle $\theta$, can be estimated by projection along the line of sight:
\begin{equation}
  \label{eq:crosscorr_a}
  w_{\rm sp}(\theta, z) = b_{\rm s}(\theta, z) \, \int_0^\infty \mathrm{d} z^\prime \, n_{\rm p}(z^\prime) \, b_{\rm p}(\theta, z^\prime) \, \xi\left[R(\theta, z, z^\prime), z\right] ,
\end{equation}
where $n_{\rm p}(z)$ is the redshift probability distribution of the target sample and $b_{\rm s}(\theta, z)$ and $b_{\rm p}(\theta, z)$ are terms for the scale dependent redshift evolution of the linear galaxy bias in both samples. Finally, $\xi(R, z)$ is the matter auto-correlation function at redshift $z$ and comoving, 3-dimensional separation
\begin{equation}
  R(\theta, z, z^\prime) = \sqrt{\left[\chi(z) - \chi(z^\prime)\right]^2 + \left[f_K(z^\prime) \, \theta\right]^2} \, .
\end{equation}
Here, $\chi(z)$ is defined as the radial comoving distance and $f_{\rm K}(z)$ as the comoving angular diameter distance to a given redshift $z$.

In practise we compute the cross-correlation in narrow bins of redshift of which each has a width of $\Delta z$ (this bin width can vary with redshift). In the following we will make a series of assumptions that allow us to simplify Eq.~(\ref{eq:crosscorr_a}) significantly. First, we assume that the redshift distributions and bias evolution terms $n_{\rm s}(z)$, $n_{\rm p}(z)$, $b_{\rm s}(z)$ and $b_{\rm p}(z)$ are constant over the interval of each redshift bin. In presence of significant sample variance, strongly varying sample selections, or insufficiently fine redshift binning, this assumption is likely violated, potentially causing biases in the recovered redshift distribution $n_{\rm p}(z)$. Secondly, we assume that the redshift bins have sufficient radial extent such that neighbouring bins are uncorrelated. This allows us to reduce the integration limits in Eq.~(\ref{eq:crosscorr_a}) to a single bin and combined, these two assumptions yield
\begin{equation}
  \label{eq:crosscorr_b}
  w_{\rm sp}(\theta, z) = n_{\rm p}(z) \, b_{\rm s}(\theta, z) \, b_{\rm p}(\theta, z) \, \int_{z}^{z + \Delta z} \mathrm{d} z^\prime \, \xi\left[R(\theta, z, z^\prime), z\right] .
\end{equation}
The remaining integral is then simply the angular matter auto-correlation function of that particular bin, $w_{\rm mm}(\theta, z)$.
Finally, we express the angular separation $\theta$ in terms of the projected physical scale
$r = \theta\,\chi / (1 + z)$ at given redshift $z$ using the flat-sky and small-angle approximations:
\begin{equation}
  \label{eq:crosscorr_c}
  w_{\rm sp}(r, z) = n_{\rm p}(z) \, b_{\rm p}(r, z) \, b_{\rm s}(r, z) \, w_{\rm mm}(r, z).
\end{equation}

In a similar manner we can derive terms for the reference and target sample angular autocorrelation functions. Analogous to Eq.~(\ref{eq:crosscorr_a}) we define
\begin{equation}
  \label{eq:autocorr_a}
  w_{\rm ss}(\theta, z) = b_{\rm s}(\theta, z) \, \int_{z}^{z + \Delta z} \mathrm{d} z^\prime \, n_{\rm s}(z^\prime) \, b_{\rm s}(\theta, z^\prime) \, \xi\left[R(\theta, z, z^\prime), z\right].
\end{equation}
Applying the same assumptions as in Eq.~(\ref{eq:crosscorr_b}) yields
\begin{equation}
  w_{\rm ss}(\theta, z) = \frac{b_{\rm s}^2(\theta, z)}{\Delta z} \, \int_{z}^{z + \Delta z} \mathrm{d} z^\prime \, \xi\left[R(\theta, z, z^\prime), z\right] ,
\end{equation}
where we have substituted $n_{\rm s}(z) \equiv 1 / \Delta z$ since the redshift distribution does not vary over $\Delta z$.
Again, we identify the integral as the angular matter auto-correlation function $w_{\rm mm}(\theta, z)$ of that particular bin and rearrange to obtain an expression for the reference sample bias evolution
\begin{equation}
  \label{eq:bias_evo}
  b_{\rm s}(\theta, z) = \sqrt{\Delta z \, \frac{w_{\rm ss}(\theta, z)}{w_{\rm mm}(\theta, z)}} \, .
\end{equation}
By repeating this approach we obtain an analogous term for the target sample bias evolution (substituting ${\rm s \rightarrow p}$ in Eqs.~\ref{eq:autocorr_a}-\ref{eq:bias_evo}).

After expressing the angles in projected physical separation we can express the bias terms in equation~(\ref{eq:crosscorr_c}) by the sample autocorrelation functions and solve for our redshift distribution of interest:  
\begin{equation}
  \label{eq:crosscorr_d}
  n_{\rm p}(z) = \frac{w_{\rm sp}(r, z)}{\sqrt{\Delta z^2 \,w_{\rm ss}(r, z) \, w_{\rm pp}(r, z)}} \, .
\end{equation}
In summary, it is possible to estimate an unknown redshift distribution $n_{\rm p}(z)$ by measuring the 
cross-correlation of $p$ with our tracer sample of known redshift $s$. However, this redshift estimate is degenerate
with the redshift evolution of the bias factors, $b_{\rm s}(z)$ and $b_{\rm p}(z)$, which can be, in principle, measured
through the individual sample auto-correlation functions $w_{\rm ss}$ and $w_{\rm pp}$.

The simple relation, presented in Eq.~(\ref{eq:crosscorr_d}), requires
some non-trivial assumption, such as linear galaxy bias, which is typically violated on small scales, and parameterising
the redshift distributions and bias terms through step functions. Depending on the chosen redshift binning and the clustering
of the reference and target sample, this can introduce significant systematic errors. Furthermore, it is very challenging
to correct for the target sample bias, since measuring $w_{\rm pp}$ requires binning the target sample into the same narrow
redshift bins $\Delta z$ that we use to slice the reference sample. This would require accurate redshift point estimates for all
galaxies in the target sample. We give a brief overview of the most common literature approaches to bias mitigation below.

\subsection{Cross-Correlation Methods and Bias Mitigation} \label{sec:cc_literature}

\citet{Newman08} and \citet{Matthews10} parameterise galaxy bias by modelling the correlation functions with power
laws: 
\begin{equation}
  \label{eq:corr_power}
  \xi(r,z) = \left(\frac{r}{r_0(z)}\right)^{-\gamma(z)},
\end{equation}
where $r_0$ is the correlation length, and $\gamma$ defines the shape of the correlation
function. Assuming linear biasing, the cross-correlation can be written as $\xi_{\rm sp} = \sqrt{\xi_{\rm ss} \,
\xi_{\rm pp}}$ and $r_{\rm 0,sp}$ and $\gamma_{\rm sp}$ can be calculated from the parameters of the auto-correlation
functions. \citet{Newman08} obtains $r_{\rm 0,ss}(z)$ and $\gamma_{\rm ss}(z)$ by fitting the reference sample's
auto-correlation function, measured in bins of redshift, and an average value for $\gamma_{\rm pp}$ by fitting the
angular auto-correlation function of the target sample. Since $r_{\rm 0,pp}$ cannot be measured without redshift
information, the authors assume that it is constant which allows to break the degeneracy of the redshift distribution
and the galaxy bias. They apply an iterative approach to obtain an estimate for $r_{\rm 0,pp}$ averaged over the redshift baseline of the target sample:
\begin{enumerate}
  \item Make an initial guess for $r_{\rm 0,pp}$ and compute $r_{\rm 0,sp}(z)$ and $\gamma_{\rm sp}(z)$.
  \item \label{enum:nz} Estimate the redshift distribution $n_{\rm p}(z)$ by fitting the measured cross-correlation with the power-law model.
  \item \label{enum:deproj} De-project the angular auto-correlation using Eq.~(\ref{eq:crosscorr_b}) and the redshift
  distribution from step \ref{enum:nz}, and thereby obtain a new guess for $r_{\rm 0,pp}$.
  \item Repeat steps \ref{enum:nz} and \ref{enum:deproj} until convergence is reached.
\end{enumerate}
\citet{Newman08} restricts the cross-correlation measurements to scales of $2 < r < 10 \si{Mpc}$ to avoid the highly non-linear
biasing regime where the assumption $\xi_{\rm sp} = \sqrt{\xi_{\rm ss} \, \xi_{\rm pp}}$ might no longer be valid.

\citet{McQuinn13} and an extension by \citet{Johnson17} build on \citet{Newman08}'s approach and construct an
estimator that optimally weights the correlation scales to improve the signal-to-noise ratio of the recovered
clustering redshift distribution.

\citet{Menard13} and \citet{Schmidt13} demonstrate that using small scales for the correlation measurements is
extremely valuable, as these scales carry the strongest correlation signal. They conclude that the systematic errors
introduced by violating the assumption of linear, deterministic bias are outweighed by an improved signal-to-noise ratio when measuring
on scales $r < \SI{1}{Mpc}$.  Furthermore, they suggest using a single-bin correlation measurement:
\begin{equation}
  \label{eq:weighted_coor}
  \bar w_{\rm sp}(z) = \int_{r_{\rm min}}^{r_{\rm max}} \mathrm{d}{r} \, W(r) \, w_{\rm sp}(r,z) \, ,
\end{equation}
where $W(r) \propto r^\beta$ is a weight function. For $\beta=-1$ this amounts to weighting galaxy pairs by
their inverse separation distance, which further increases the signal-to-noise ratio and the sensitivity to galaxy bias by up-weighting the smallest scales.

The authors also suggest two methods of bias mitigation for the reference and target samples.
First, the reference sample auto-correlation function yields an estimate of the bias
evolution with redshift, when measured on the same scales and with the same weighting as the cross-correlation
function (see Eq.~\ref{eq:crosscorr_d}):
\begin{equation}
  \label{eq:spec_bias}
  \tilde n_{\rm p}(z) = \frac{\bar w_{\rm sp}(z)}{\sqrt{\Delta z \, \bar w_{\rm ss}(z)}} = n_{\rm p}(z) \sqrt{\Delta z \, \bar w_{\rm pp}(z)} \, ,
\end{equation}
where the barred correlation functions are scale-weighted according to Eq.~(\ref{eq:weighted_coor}).
Here we have defined $\tilde n_{\rm p}(z)$ as clustering redshift distribution corrected for the reference sample bias. Secondly,
they note the correlation between the width of the target sample redshift distribution and its sensitivity to the
galaxy bias redshift evolution. By pre-selecting narrow redshift bins (e.g. using magnitudes, colours or photometric redshifts)
the impact of the redshift-evolution of the bias can be reduced. Additionally, \citet{Newman08}'s iterative bias technique can be applied to
each bin individually to recover the step-wise redshift evolution of the bias.

Finally, \citet{Davis18} parameterise the bias of the target sample through a simple power-law:
\begin{equation}
  \label{eq:bias_model}
  \mathcal{B}_\alpha(z) = (1+z)^\alpha \propto \sqrt{\Delta z \, \bar w_{\rm pp}(z)} \, . 
\end{equation}
The normalisation of this parameterisation is arbitrary; it is degenerate with the normalisation of the
resulting redshift distribution.

\subsection{Bias mitigation with self-consistency} \label{sec:bias_fit}

In \citet{Hildebrandt20a} we adopted a method to constrain the bias of the target sample using an analytical model
(e.g. Eq.~\ref{eq:bias_model}), which is based on ideas developed in \citet{Schmidt13}, \citet{Menard13}, and \citet{Morrision17}.
The method leverages the response of clustering redshifts to the target sample bias as a function of 
the target sample redshift distribution width.

\subsubsection{Method}

\begin{figure}[t]
  \centering
  \includegraphics{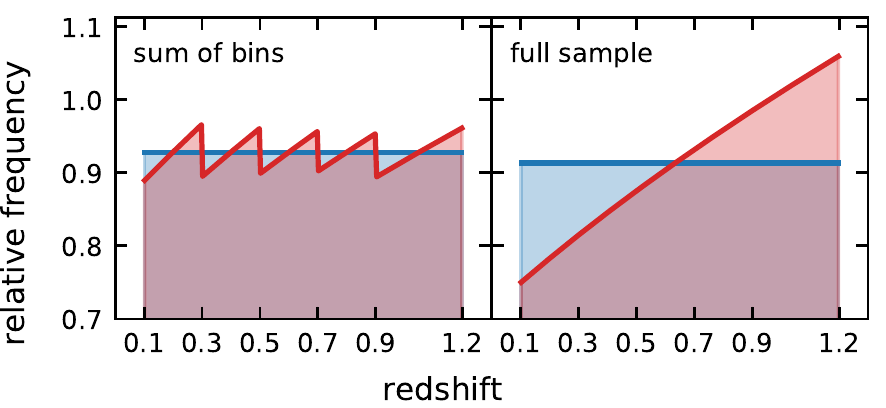}
  \caption{
    Toy-model showing the impact of the bias evolution on clustering redshifts for wide redshift distribution compared to measurements on narrow redshift bins.}
  \label{fig:fit_dummy}
\end{figure}

We split the target sample into
$N_{\rm bin}$ (preferentially narrow) bins using a secondary redshift indicator, such as photometric redshifts.
Then we measure clustering redshifts $\tilde n_{{\rm p}, j}(z)$ for each of these bins.
If the bias evolves with redshift, the sum of these binned measurements must differ from a clustering redshift
measurement $\tilde n_{\rm p,tot}(z)$ (Eq.~\ref{eq:spec_bias}) over the full target sample:
\begin{equation}
  \label{eq:sum_inequality}
  \sum_{j=1}^{N_{\rm bin}} \mathcal{W}_j \, \tilde n_{{\rm p},j}(z) \neq \tilde n_{\rm p,tot}(z) \, ,
\end{equation}
where $\mathcal{W}_j$ is the total weight of the $j$-th tomographic bin, obtained by summing over the individual weights of all galaxies contained in that bin.
This is because each of the $\tilde n_{{\rm p}, j}(z)$ is normalised individually and this normalisation factor depends
on the mean bias amplitude per bin and allows, in principle, to break the degeneracy between redshift and bias evolution.
A toy-model demonstration of this effect can be seen in Fig.~\ref{fig:fit_dummy}. We assume a flat redshift distribution in all bins, shown in blue, and a redshift evolution of the sample bias of $\mathcal{B}_{\alpha}(z) \propto (1+z)^\alpha$, where we set $\alpha=0.5$. This results in the recovered, biased clustering redshift distributions shown in red. The left- and right-hand side of this figure correspond to the left- and right-hand side terms of Eq.~(\ref{eq:sum_inequality}). Whereas the full sample has the complete bias evolution imprinted, the normalisation of each redshift bin leads to a sawtooth-shaped redshift distribution.

These differences between the full sample redshift distribution and the sum of the redshift bins allows us to constrain the bias evolution, if we have a sufficiently accurate model. We can estimate the model parameter $\alpha$ of the bias model $\mathcal{B}_{\alpha}(z)$ by minimising
\begin{equation}
  \label{eq:fit_residual}
  \Delta(\alpha) = \int \mathrm{d}z \, \left[ \operatorname{norm}\left(\frac{\tilde n_{\rm p,tot}(z)}{\mathcal{B}_\alpha(z)}\right) - \sum_{j=1}^{N_{\rm bin}} \mathcal{W}_j \, \operatorname{norm}\left(\frac{\tilde n_{{\rm p},j}(z)}{\mathcal{B}_\alpha(z)}\right) \right]^2 ,
\end{equation}
where $\operatorname{norm}[f(z)] = f(z) \, \big/ \, \int_0^\infty \mathrm{d}{z^\prime} \, f(z^\prime)$.

The advantage of this approach is that it allows us to correct for any combination of biases in clustering redshifts, $\bar b_{\rm p}(z)$, $\bar b_{\rm s}(z)$ or the combination of both. The bias model, however, must be fairly simplistic since the amount of information that can be extracted from Eq.~(\ref{eq:fit_residual}) is small and depends on total number of bins and the degree of overlap of their respective redshift distributions. The greater the overlap between the redshift bins is, the stronger is the correlation between their redshift distributions and the harder it is to constrain the bias evolution. Furthermore, this approach assumes that all bins follow the same universal bias evolution. This is not true in general, since any redshift binning will select a different population of galaxies which cluster differently. This effect is small for the tomographic redshift bins of the KV450 mock galaxy sample, but may be larger for more complex sample selections.
Finally, we note that this mitigation approach can be applied to a variety of the existing clustering-redshift methods (see overview in Sect.~\ref{sec:cc_literature}) as a post-processing step.

We will call this mitigation approach `self-consistent bias mitigation' (SBM) in the following.

\subsubsection{Tests on mock data} \label{sec:bias_fit_result}

\begin{figure}[t]
  \centering
  \includegraphics[width=\columnwidth]{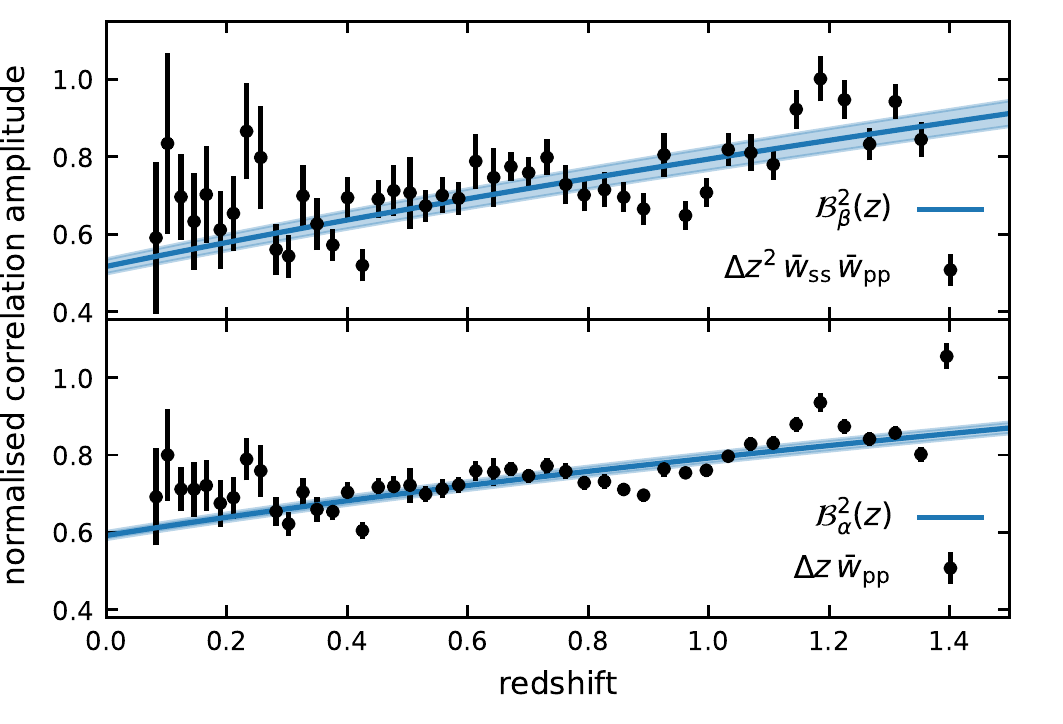}
  \caption{
    Bias model (blue lines; corresponding to the bold-face numbers in Table~\ref{tab:alphas}) fitted
    directly to the unaccounted bias terms of the raw (black data points; top panel) and the reference
    sample bias corrected clustering redshifts (black data points; bottom panel) for the idealised
    setup. The correlation amplitudes are re-normalised.}
  \label{fig:alphas_true}
\end{figure}

\begin{table}[t]
  \centering
  \caption{
    Best-fit values for the bias model parameter $\beta$ (modelling the reference and target sample bias), $\alpha$ (modelling only the target sample bias), and $\gamma$ (null test). The top group summarises the results obtained using the SBM and the bottom group results from direct fits to the measured correlation function terms.}
  \label{tab:alphas}
  \resizebox{\columnwidth}{!}{
    \begin{tabular}{ll|ccc}
  \hline\hline
  Method & Setup & $\beta$ & $\alpha$ & $\gamma$ \\
  \hline
  \multirow{2}{*}{SBM} & idealised & $-0.03 \pm 0.11$ & $-0.09 \pm 0.10$ & $-0.26 \pm 0.10$ \\
   & realistic & $\hphantom{-}0.50 \pm 1.05$ & $\hphantom{-}0.88 \pm 1.05$ & $\hphantom{-}0.82 \pm 1.04$ \\
  \hline
  \multirow{2}{*}{direct fit} & idealised & $\hphantom{-}0.31 \pm 0.03^{(\star)}$ & $\hphantom{-}0.21 \pm 0.01^{(\star)}$ & $\hphantom{-}$--- \\
   & realistic & $-0.06 \pm 0.04$ & $\hphantom{-}0.18 \pm 0.02$ & $\hphantom{-}$--- \\
  \hline
\end{tabular}
}
  \tablefoot{
    The top group summarises the fit results from the SBM, the bottom group from direct fits
    to the auto-correlation function terms. The two values marked by $(\star)$ correspond to the
    blue model fits in Fig.~\ref{fig:alphas_true}.}
\end{table}

We apply the SBM to both, clustering redshifts obtained from the idealised reference sample as well as from the compilation of realistic reference samples, to assess whether it is able to recover the bias evolution terms correctly.
In either case we expect the SBM best fit to match $\sqrt{\Delta z \, \bar w_{\rm pp}(z)}$ (see Eq.~\ref{eq:spec_bias}) when applied to $\tilde{n}_{\rm p}(z)$, where we have already corrected the reference sample bias by measuring its autocorrelation function. We conduct additional tests, namely fitting the raw cross-correlations $\bar w_{\rm sp}(z)$ with
\begin{equation}
  \label{eq:beta_model}
  \mathcal{B}_\beta(z) = (1+z)^\beta \propto \sqrt{\Delta z^2 \, \bar w_{\rm ss}(z) \, \bar w_{\rm pp}(z)}
\end{equation}
and fitting the fully bias-corrected clustering redshifts $n_{\rm p}(z)$ (Eq.~\ref{eq:crosscorr_d}) with $\mathcal{B}_\gamma(z) = (1+z)^\gamma$. The latter serves as a null test, since there should be no residual bias evolution, hence $\mathcal{B}_\gamma(z) \approx 1$ and $\gamma \approx 0$.

We validate these SBM parameter estimates by directly fitting the bias model to the autocorrelation terms, i.e. fitting the right hand side terms with the left hand side models of Eqs.~(\ref{eq:spec_bias}) and~(\ref{eq:beta_model}). These autocorrelation terms and direct fits are presented in Fig.~\ref{fig:alphas_true}.
The correlation measurements reveal that the bias evolution is small for all the mock data samples. The galaxy bias of the target sample increases by $\approx\SI{25}{\percent}$
over the MICE2 redshift baseline. Furthermore, the direct fitting
shows that the power-law bias model (Eq.~\ref{eq:bias_model}) is able
to recover the global trend of the bias evolution, despite having only a single free parameter. The
best-fit values for $\alpha$, $\beta$ and $\gamma$ obtained from both, the SBM and the direct auto-correlation function
fits, are summarised in Table~\ref{tab:alphas}.

The results we obtain from the idealised mock sample using the SBM are not in agreement with
the direct fits and instead predict negative values in all three cases. The values themselves are relatively
small though because the idealised sample is purely magnitude limited and hence should not show a very
strong bias evolution. This example illustrates the limitations of the SBM. The impact of this
on the mean redshifts is discussed in Sect.~\ref{sec:bias_method_results}.

In case of the realistic mocks we find that $\alpha$,
determined via the SBM, is of order unity but, due to the large parameter uncertainty, consistent
with the value expected from the idealised setup.
We note that the uncertainties reported in Table~\ref{tab:alphas} are significantly larger than those reported in \citet{Hildebrandt20a}. This results from \thewizz{} error estimates adopted by \citet{Hildebrandt20a}, which we find underestimate the true uncertainty on the measured correlation functions at high redshifts. Our revised pipeline \yaw{} (see Sect.~\ref{sec:cc_code} below) provides accurate error estimates that serve to increase the overall uncertainty on $\alpha$.

\subsection{Implementation} \label{sec:cc_code}

We have used clustering redshifts to obtain the redshift distributions of the KiDS-450 and the KiDS+VIKING-450 datasets in previous works \citep{Hildebrandt17,Hildebrandt20a}. The major difference to the latter is that we are using an improved implementation of the \citet{Schmidt13} clustering-redshift method.

Previously, we used \thewizz{}\footnote{\url{https://github.com/morriscb/the-wizz}} \citep{Morrision17} for clustering redshifts and a \texttt{python}-binding\footnote{\url{https://github.com/jlvdb/astro-stomp3}} of the spherical pixelation library \stomp{} \citep{Scranton02} for efficient angular correlation measurements. The issue with this approach is that \stomp{} (and therefore \thewizz{}) approximates the correlation annulus, in which the pairs are counted, by selecting entire \stomp{} pixels. Since we compute correlations on constant physical scales, the area of the annulus may change discretely with position and redshift, depending on the \stomp{} pixel resolution and the angular diameter corresponding to the given projected physical separation. This can cause biases in the recovered redshift distribution.
During the development stage of this project, the latest version of \thewizz{}, which no longer depends on \stomp{}, was not yet available.
Therefore we implemented a simplified version of \thewizz{}, called \yaw{}\footnote{\url{https://github.com/jlvdb/yet\_another\_wizz}} to circumvent the pixelation problems. Similar to \thewizz{}, this code measures the correlation on a single annulus of fixed projected physical separation $r_{\rm p}$, weighted by the inverse distance of the pairs. Even though the calculation of $r_{\rm p}(z)$ requires assuming a cosmological model, the results are only very weakly dependent on the choice of cosmological parameters. The inverse distance weight significantly increases the signal-to-noise ratio of the correlation amplitude \citep{Schmidt13}. Similar approaches have been used e.g. by \citet{McQuinn13} or \citet{Alarcon20} to optimise the correlation signal.

We compute the single-bin correlation amplitude (see Eq.~\ref{eq:weighted_coor}) from pair counts using the Davis-Peebles correlation estimator \citep{Davis83}
\begin{equation}
  \label{eq:estimator}
  \bar w = \frac{n_{\rm R} \int_{r_{\rm min}}^{r_{\rm max}} \mathrm{d}{r} \, W(r) \, \operatorname{DD}(r)}{n_{\rm D} \int_{r_{\rm min}}^{r_{\rm max}} \mathrm{d}{r} \, W(r) \, \operatorname{DR}(r)} - 1
\end{equation}
with an inverse distance weight $W(r) \propto r^{-1}$. $\operatorname{DD}$ is the number of (ordered) pairs\footnote{Each pair is weighted by the product of the individual weights of the partners.} between reference and target sample galaxies and $\operatorname{DR}$ between reference galaxies and target sample random points.
The ratio $n_{\rm R} / n_{\rm D}$ (sum over the galaxy weights divided by the sum over the random point weights) accounts for the average density of the data catalogue and its typically much larger random representation. In general it would be favourable to use the Landy-Szalay estimator \citep{Landy93} instead of the Davis-Peebles estimator, albeit this would come at the cost of increased computational complexity. We find that the difference between the estimators in our analysis is negligible compared to other systematic errors, such as the evolving target sample bias. Finally, we choose to measure all angular correlations on scales from $r_{\rm min} = \SI{100}{kpc}$ to $r_{\rm max} = \SI{1000}{kpc}$. This range is a good trade-off between high signal-to-noise and measuring on scales with highly non-linear biasing.

The whole clustering-redshift pipeline as well as our bootstrap resampling-based method for combining cross-correlation measurements from different reference samples and covariance estimation is described in detail in Appendix~\ref{sec:pipeline}.


\section{Results} \label{sec:results}

In this section we report the results of applying our method to the KV450 simulations. We focus on the effects
of galaxy bias, the choice of measurement scales, and the effect of lensing magnification.

\subsection{Impact of galaxy bias} \label{sec:bias_method_results}

There are different degrees of bias correction that can be applied before getting a redshift estimate $n_{\rm CC}(z)$
from cross-correlation measurements:
\begin{enumerate}
  \item The raw cross-correlation with no correction ($n_{\rm CC} = \bar w_{\rm sp}$),
  \item Mitigating the reference sample bias using its auto-correlation function ($n_{\rm CC} = \tilde n_{\rm p}$, Eq.~\ref{eq:spec_bias}),
  \item Additionally mitigating the target sample bias with the SBM 
  ($n_{\rm CC} = \tilde n_{\rm p} / \mathcal{B}_\alpha$), and
  \item Mitigating the target sample bias using its auto-correlation function ($n_{\rm CC} = n_{\rm p}$, Eq.~\ref{eq:crosscorr_d}, only
  possible on mock data using the true redshifts).
\end{enumerate}
It is difficult to directly compare each of these four redshift estimates. Due to negative correlation amplitudes, originating from systematic effects and statistical noise, it is dangerous to interpret cross-correlation-derived redshift estimates directly as probability distributions. This prevents us from using them directly in cosmological applications such as cosmic shear, unless we model the redshift distributions such that negative amplitudes are mitigated.

We implement this modelling by adopting the true target sample redshift distribution $p(z)$ as a model and fit it to the clustering redshift estimate $n_{\rm CC}(z)$, allowing a free normalisation amplitude $A$ and a free shift parameter $\Delta z$. We minimise
\begin{equation}
  \label{eq:shift_fit}
  \chi^2 = \left[ n_{\rm CC}(z) - A \, p(z + \Delta z) \right]^T C^{-1} \left[ n_{\rm CC}(z) - A \, p(z + \Delta z) \right] \, ,
\end{equation}
where we shift $p(z)$ according to $\Delta z$ and apply the binning of the cross-correlation to the model redshift distribution. We perform these fits jointly for all tomographic bins\footnote{See Appendix~\ref{sec:cc_code} for a summary of the covariance estimation recipe.} to capture the full correlation of the shift parameters $\Delta z_i$.
These shift parameters $\Delta z_i \approx \langle z_{\rm CC} \rangle_i - \langle z_{\rm true} \rangle_i$ (presented in Sect.~\ref{sec:true_shifts}) give us a direct estimate of the systematic shifts in the four different redshift estimates listed above, which may originate from evolving galaxy bias or a breakdown of the assumption of the cross-correlation formalism (see Sect.~\ref{sec:formalism}). Albeit, one has to keep in mind that, by design, this shift-fitting approach is mostly sensitive to the overall shape of the redshift distribution rather than individual outliers or noise fluctuations at the tails of the distribution.

\begin{table*}[t]
  \centering
  \caption{
    Shift fit parameters for different bias correction methods for clustering redshifts obtained using the idealised and realistic spectroscopic mock samples using the true redshift distributions as fit model.}
  \label{tab:shifts_true}
  \renewcommand{\arraystretch}{1.3}
  \begin{tabular}{ll|ccccccc}
  \hline\hline
  Setup & $n(z)$-type & $100\times \Delta z_1$ & $100\times \Delta z_2$ & $100\times \Delta z_3$ & $100\times \Delta z_4$ & $100\times \Delta z_5$ & $\chi^2$ & $n_{\rm dof}$ \\
  \hline
  \multirow{4}{*}{idealised} & $\bar w_{\rm sp}$ & $-0.08_{-0.13}^{+0.11}$ & $-0.12_{-0.11}^{+0.09}$ & $-0.40_{-0.08}^{+0.12}$ & $-0.79_{-0.20}^{+0.13}$ & $-0.07_{-0.18}^{+0.20}$ & \hphantom{0}263.0 & 215 \\
   & $\tilde n_{\rm p}$ & $-0.22_{-0.14}^{+0.08}$ & $-0.15_{-0.10}^{+0.10}$ & $-0.51_{-0.12}^{+0.12}$ & $-0.80_{-0.18}^{+0.12}$ & $-0.26_{-0.19}^{+0.16}$ & \hphantom{0}261.5 & 215 \\
   & $\tilde n_{\rm p} / \mathcal{B}_\alpha$ & $-0.15_{-0.12}^{+0.12}$ & $-0.14_{-0.10}^{+0.09}$ & $-0.45_{-0.10}^{+0.13}$ & $-0.70_{-0.14}^{+0.16}$ & $-0.19_{-0.23}^{+0.21}$ & \hphantom{0}249.5 & 215 \\
   & $n_{\rm p}$ & $-0.24_{-0.11}^{+0.12}$ & $-0.16_{-0.09}^{+0.10}$ & $-0.68_{-0.13}^{+0.10}$ & $-0.82_{-0.17}^{+0.10}$ & $-0.76_{-0.33}^{+0.41}$ & \hphantom{0}282.0 & 215 \\
  \hline
  \multirow{4}{*}{realistic} & $\bar w_{\rm sp}$ & $-0.32_{-2.10}^{+0.47}$ & $\hphantom{-}0.74_{-0.17}^{+0.19}$ & $\hphantom{-}0.01_{-0.39}^{+0.20}$ & $-0.41_{-0.49}^{+0.44}$ & $-2.55_{-2.89}^{+2.27}$ & \hphantom{0}234.6 & 100 \\
   & $\tilde n_{\rm p}$ & $-0.44_{-0.48}^{+0.43}$ & $\hphantom{-}0.31_{-0.19}^{+0.22}$ & $-0.36_{-0.24}^{+0.19}$ & $\hphantom{-}0.60_{-0.58}^{+0.64}$ & $\hphantom{-}0.32_{-2.40}^{+0.49}$ & \hphantom{0}157.3 & 100 \\
   & $\tilde n_{\rm p} / \mathcal{B}_\alpha$ & $\hphantom{-}0.83_{-0.93}^{+0.37}$ & $\hphantom{-}0.20_{-0.60}^{+0.41}$ & $-0.17_{-0.60}^{+0.54}$ & $-1.55_{-0.71}^{+0.38}$ & $-0.93_{-1.18}^{+4.00}$ & \hphantom{00}93.7 & 100 \\
   & $n_{\rm p}$ & $-0.88_{-0.33}^{+0.25}$ & $\hphantom{-}0.38_{-0.14}^{+0.20}$ & $-0.46_{-0.26}^{+0.16}$ & $\hphantom{-}0.61_{-0.56}^{+0.65}$ & $\hphantom{-}0.30_{-0.45}^{+0.87}$ & \hphantom{0}169.2 & 100 \\
  \hline
\end{tabular}

  \renewcommand{\arraystretch}{1.0}
\end{table*}

\begin{figure*}[t]
  \centering
  \hfill
  \includegraphics[width=0.95\columnwidth]{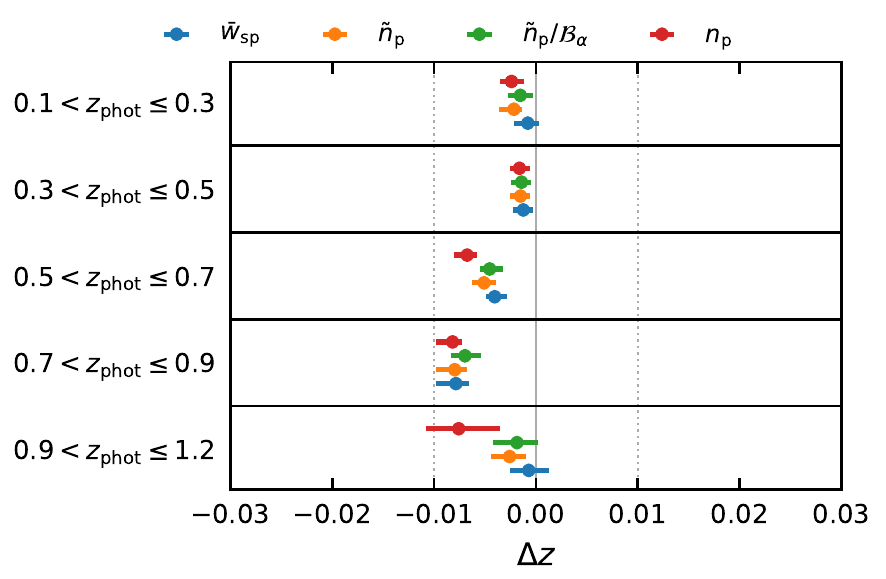}
  \hfill
  \includegraphics[width=0.95\columnwidth]{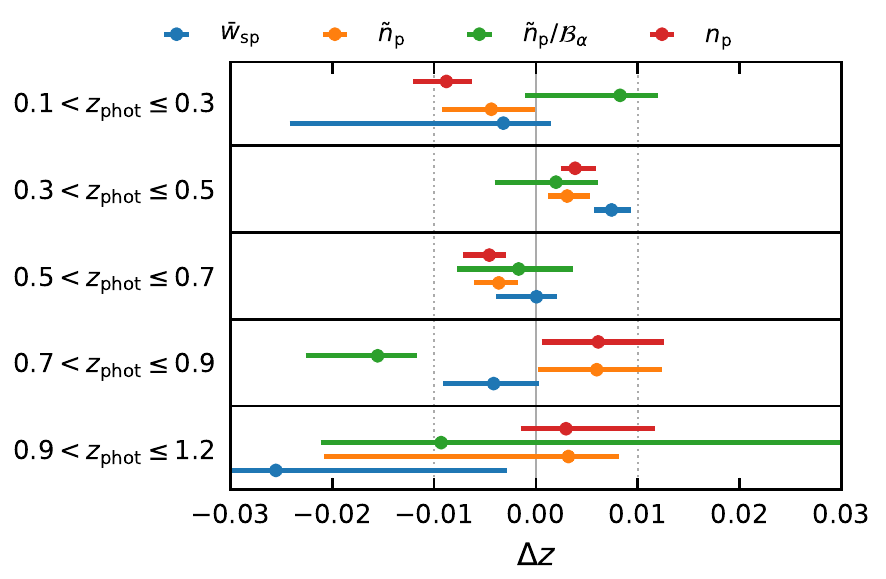}
  \caption{
    Visualisation of the shift parameters $\Delta z_i$ from fitting the true redshift distributions
    to the clustering redshifts obtained using the idealised (left side) and the realistic (right side) spectroscopic mock samples. The colours
    indicate different bias correction methods applied: the raw cross-correlation (blue), reference sample bias corrected (orange), additionally the target sample bias corrected using the SBM (green) and the target sample bias corrected using the sample autocorrelation function (red).}
  \label{fig:true_shifts}
\end{figure*}

In practical applications we do not have access to the true target sample redshift distributions. We therefore repeat the procedure with redshift distributions constructed using the direct calibration method \citep[DIR,][]{Lima08,Hildebrandt20a} as fit model (presented in Sect.~\ref{sec:DIR_shifts}). In short, these DIR redshift distributions are obtained by re-weighting a spectroscopic calibration sample such that it has the same properties in colour and magnitude space as the target sample (the KV450 cosmic shear sample) and are therefore fundamentally different from clustering redshifts. By using photometrically determined redshift estimates as a fit model, our approach can be considered similar to the DES redshift calibration, presented in \citet{Gatti18}. We however approach the problem from a different perspective: Instead of using the clustering redshifts to calibrate the DIR redshift distributions, we utilise the DIR to interpret the clustering redshifts. One can easily construct a variety of more sophisticated and hence more complex models than simply shifting an existing redshift estimate to interpret clustering redshifts. However, fitting these models requires a better understanding of the data covariance, in particular at high redshifts, which are dominated by a small number of deep spectroscopic fields.

\begin{figure*}[t]
  \centering
  \includegraphics[width=\textwidth]{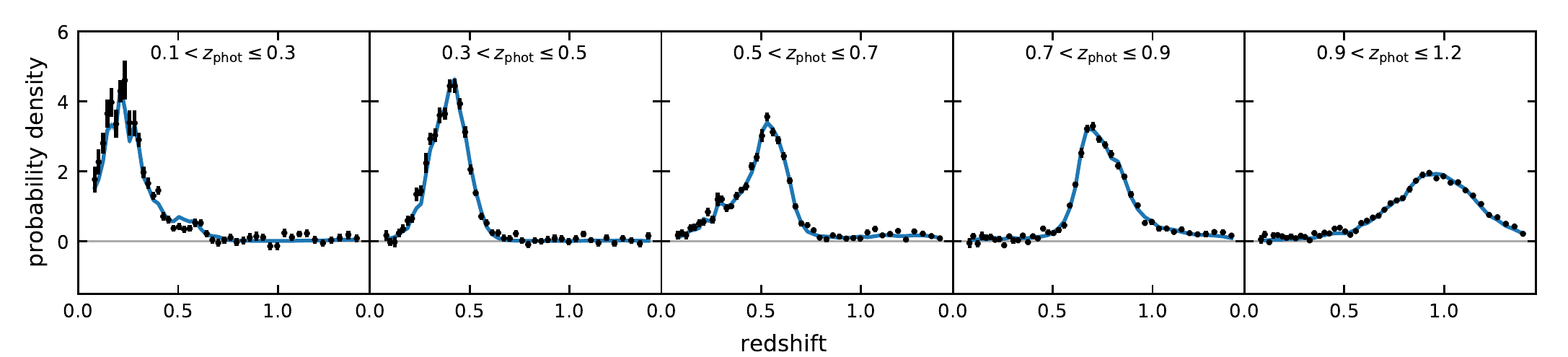}
  \caption{
    Plot of the reference sample bias corrected clustering redshifts (black data points) fitted with the shifted true redshift distributions (blue, re-binned to the 45 data points) for all tomographic bins of the idealised mocks.}
  \label{fig:true_idealized_model_plot}
\end{figure*}

\begin{figure*}[t]
  \centering
  \includegraphics[width=\textwidth]{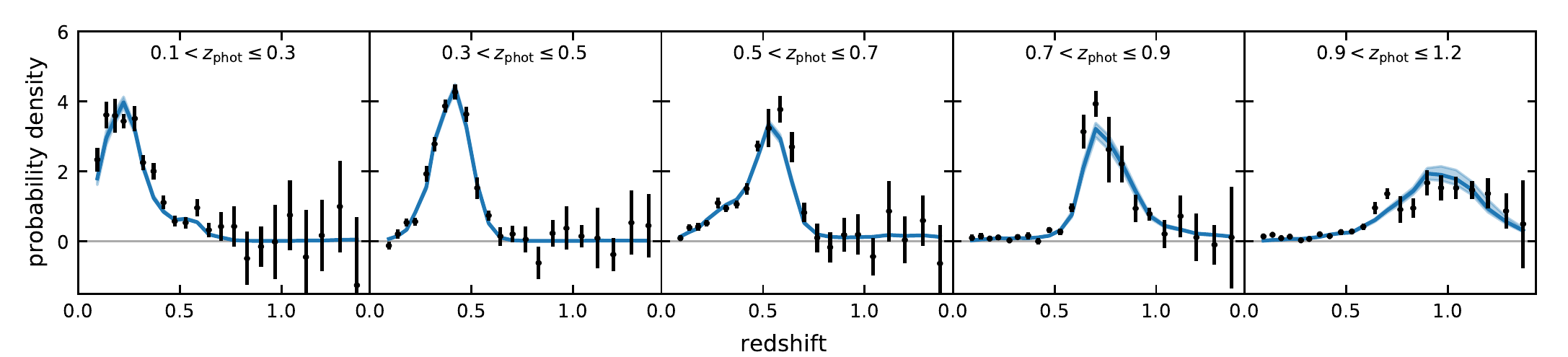}
  \caption{
    Plot of the reference sample bias corrected clustering redshifts (black data points) fitted with the shifted DIR redshift distributions (blue, re-binned to the 22 data points) for all tomographic bins of the realistic mocks.}
  \label{fig:true_realistic_model_plot}
\end{figure*}

\begin{table*}[t]
  \centering
  \caption{
    Shift fit parameters for different bias correction methods for clustering redshifts obtained using the idealised and realistic spectroscopic mock samples using the DIR redshift distributions as fit model.}
  \label{tab:shifts_DIR}
  \renewcommand{\arraystretch}{1.3}
  \begin{tabular}{ll|ccccccc}
  \hline\hline
  Setup & $n(z)$-type & $100\times \Delta z_1$ & $100\times \Delta z_2$ & $100\times \Delta z_3$ & $100\times \Delta z_4$ & $100\times \Delta z_5$ & $\chi^2$ & $n_{\rm dof}$ \\
  \hline
  \multirow{4}{*}{idealised} & $\bar w_{\rm sp}$ & $\hphantom{-}0.98_{-0.52}^{+0.50}$ & $\hphantom{-}0.73_{-0.42}^{+0.43}$ & $\hphantom{-}3.02_{-0.56}^{+0.53}$ & $-2.06_{-0.43}^{+0.42}$ & $\hphantom{-}0.25_{-0.43}^{+0.45}$ & \hphantom{}2069.5 & 215 \\
   & $\tilde n_{\rm p}$ & $\hphantom{-}1.03_{-0.52}^{+0.50}$ & $\hphantom{-}0.77_{-0.43}^{+0.42}$ & $\hphantom{-}2.83_{-0.52}^{+0.53}$ & $-2.21_{-0.43}^{+0.42}$ & $-0.09_{-0.46}^{+0.45}$ & \hphantom{}2371.1 & 215 \\
   & $\tilde n_{\rm p} / \mathcal{B}_\alpha$ & $\hphantom{-}1.18_{-0.58}^{+0.53}$ & $\hphantom{-}0.95_{-0.42}^{+0.44}$ & $\hphantom{-}2.85_{-0.52}^{+0.52}$ & $-2.16_{-0.43}^{+0.42}$ & $-0.09_{-0.46}^{+0.45}$ & \hphantom{}2123.4 & 215 \\
   & $n_{\rm p}$ & $\hphantom{-}1.18_{-0.55}^{+0.52}$ & $\hphantom{-}0.77_{-0.43}^{+0.42}$ & $\hphantom{-}2.49_{-0.51}^{+0.53}$ & $-2.37_{-0.43}^{+0.42}$ & $-0.66_{-0.54}^{+0.47}$ & \hphantom{}2587.4 & 215 \\
  \hline
  \multirow{4}{*}{realistic} & $\bar w_{\rm sp}$ & $\hphantom{-}3.31_{-0.88}^{+0.60}$ & $\hphantom{-}2.39_{-0.40}^{+0.40}$ & $\hphantom{-}4.56_{-0.52}^{+0.51}$ & $\hphantom{-}0.10_{-0.49}^{+0.94}$ & $-1.37_{-1.61}^{+0.89}$ & \hphantom{0}635.9 & 100 \\
   & $\tilde n_{\rm p}$ & $\hphantom{-}2.24_{-0.72}^{+0.67}$ & $\hphantom{-}1.88_{-0.43}^{+0.47}$ & $\hphantom{-}3.37_{-0.69}^{+0.50}$ & $\hphantom{-}0.97_{-0.40}^{+0.44}$ & $\hphantom{-}1.31_{-0.81}^{+0.84}$ & \hphantom{0}638.8 & 100 \\
   & $\tilde n_{\rm p} / \mathcal{B}_\alpha$ & $\hphantom{-}4.46_{-1.63}^{+0.97}$ & $\hphantom{-}2.37_{-0.40}^{+0.59}$ & $\hphantom{-}4.46_{-0.90}^{+2.66}$ & $-2.64_{-0.82}^{+0.76}$ & $-0.77_{-1.87}^{+2.03}$ & \hphantom{0}203.6 & 100 \\
   & $n_{\rm p}$ & $\hphantom{-}2.15_{-0.68}^{+0.60}$ & $\hphantom{-}2.04_{-0.41}^{+0.42}$ & $\hphantom{-}3.47_{-0.61}^{+0.52}$ & $\hphantom{-}0.97_{-0.40}^{+0.46}$ & $\hphantom{-}0.70_{-0.83}^{+0.71}$ & \hphantom{0}675.3 & 100 \\
  \hline
\end{tabular}

  \renewcommand{\arraystretch}{1.0}
\end{table*}

\begin{figure*}[t]
  \centering
  \hfill
  \includegraphics[width=0.95\columnwidth]{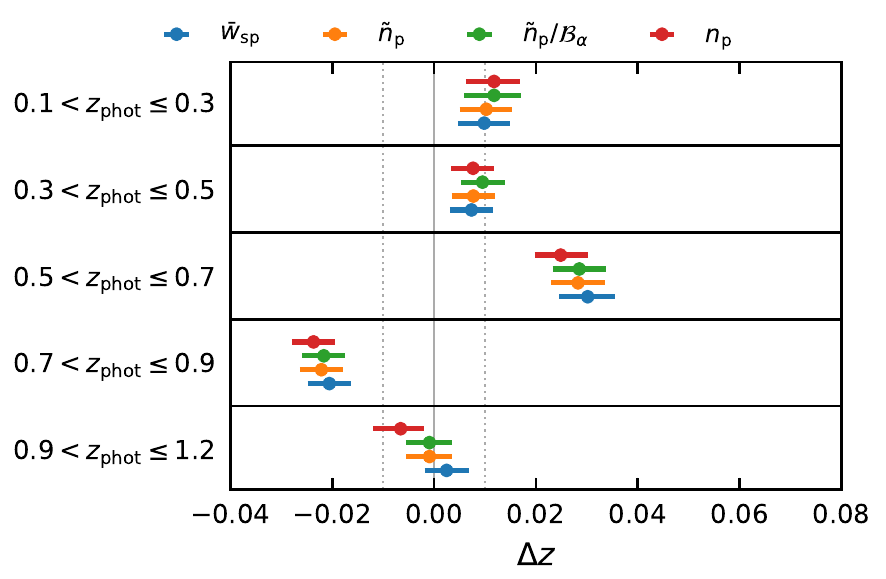}
  \hfill
  \includegraphics[width=0.95\columnwidth]{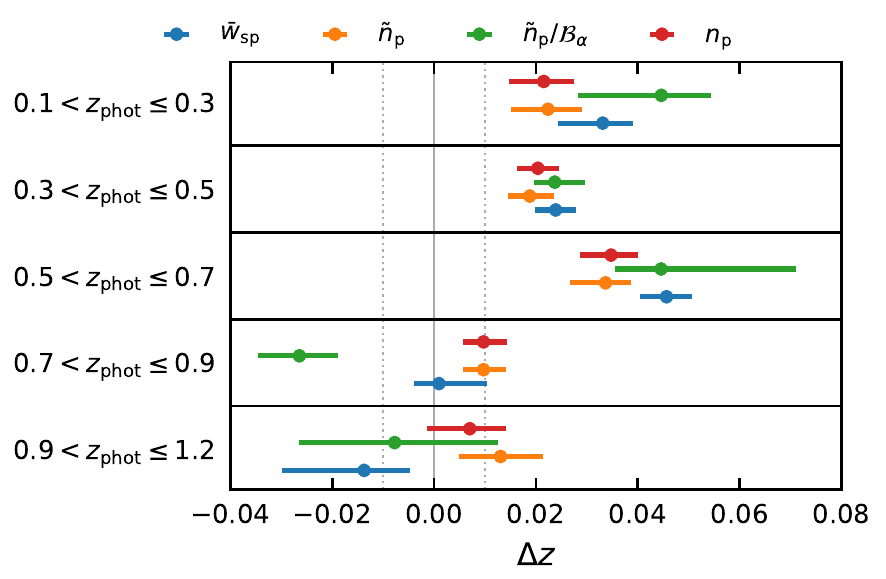}
  \caption{
    Visualisation of the shift parameters $\Delta z_i$ from fitting the DIR redshift distributions
    to the clustering redshifts obtained using the idealised (left side) and the realistic (right side) spectroscopic mock samples. The colours
    indicate different bias correction methods applied: the raw cross-correlation (blue), reference sample bias corrected (orange), additionally the target sample bias corrected using the SBM (green) and the target sample bias corrected using the sample autocorrelation function (red).}
  \label{fig:DIR_shifts}
\end{figure*}

\subsubsection{Clustering-$z$ accuracy given the true $n(z)$} \label{sec:true_shifts}

The best-fit $\Delta z$ and $\chi^2$-values for using the true redshift distributions as fit model are summarised in Table~\ref{tab:shifts_true} and visualised
in Fig.~\ref{fig:true_shifts}. Almost all shift parameters are smaller than $|\Delta z| < 0.01$, the
goal for the KiDS redshift calibration, indicating an insignificant overall bias of the recovered
redshift distributions. The clustering redshifts obtained from the idealised reference sample suggest
a slight tendency to underestimate the true redshifts, especially in the third and fourth tomographic bins.
These shifts are likely related to the finite binning of the clustering redshifts with a constant comoving
width of $\Delta \chi \approx \SI{88}{Mpc}$, which results in a different number of sampling points that
cover the peaks of each of the tomographic bins. Furthermore, the results are insensitive to bias
corrections given the small evolution of the bias with redshift of the idealised and target
samples (see Fig.~\ref{fig:alphas_true}). The clustering-$z$ data points for
$\tilde n_{\rm p}(z)$ and the best-fit model are shown in Fig.~\ref{fig:true_idealized_model_plot}. 
The signal-to-noise ratio is good enough to reveal details of the redshift distributions such as the
outlier population in the tail of the third tomographic bin.

This is no longer true when using the realistic reference sample (Fig.~\ref{fig:true_realistic_model_plot}). There is a smaller fraction of high redshift reference
galaxies, considerably increasing the uncertainty on the redshift distributions at $z > 0.7$. Still,
the overall shape of the clustering redshifts closely matches the true redshift distributions
after correcting for the reference sample bias. In the highest tomographic bin, the shift parameter
reduces significantly to be within $|\Delta z| < 0.01$ after removal of the reference sample bias via
its auto-correlation function. As for the idealised case, addressing the target sample bias does not
improve the results any further. To the contrary, the SBM bias correction increases the shifts
and their uncertainty in most cases.

\subsubsection{Clustering-$z$ accuracy given the DIR $n(z)$} \label{sec:DIR_shifts}

\begin{figure*}[t]
  \centering
  \includegraphics[width=\textwidth]{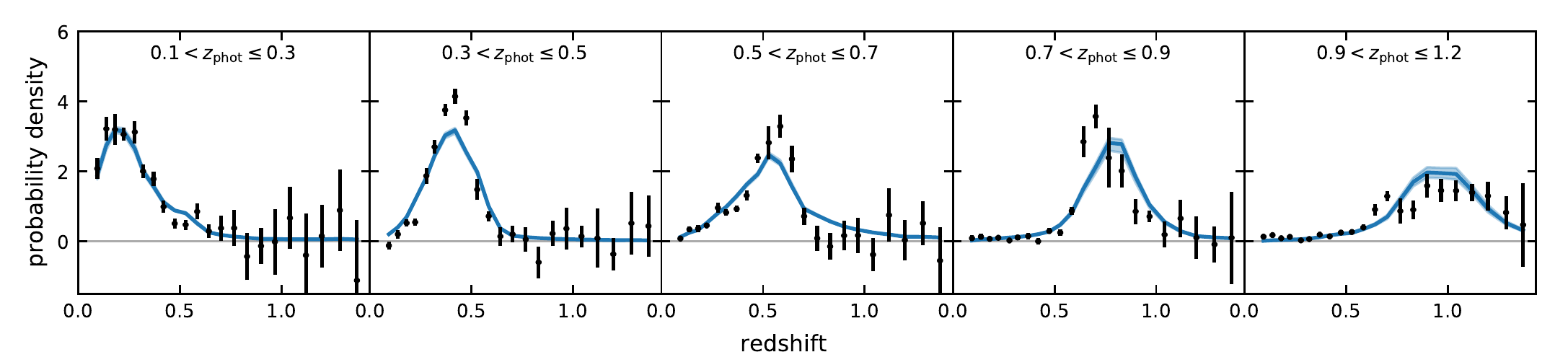}
  \caption{
    Plot of the reference sample bias corrected clustering redshifts (black data points) fitted with the shifted DIR redshift distributions (blue, re-binned to the 22 data points) for all tomographic bins of the realistic mocks.}
  \label{fig:DIR_realistic_model_plot}
\end{figure*}

\begin{table*}[t]
  \centering
  \caption{
    Shift fit parameters for two different correlation scales determined from the idealised mock data,
    once with lensing enabled and once disabled. The fits are based on $\tilde{n}_{\rm p}(z)$ using
    the true redshift distributions as model.}
  \label{tab:shifts_systematics}
  \renewcommand{\arraystretch}{1.3}
  \begin{tabular}{ll|ccccccc}
  \hline\hline
  Scale (kpc) & Lensing & $100\times \Delta z_1$ & $100\times \Delta z_2$ & $100\times \Delta z_3$ & $100\times \Delta z_4$ & $100\times \Delta z_5$ & $\chi^2$ & $n_{\rm dof}$ \\
  \hline
  \multirow{2}{*}{$100 - 1000$} & off & $-0.40_{-0.12}^{+0.12}$ & $-0.18_{-0.09}^{+0.06}$ & $-0.39_{-0.10}^{+0.10}$ & $-0.60_{-0.19}^{+0.13}$ & $\hphantom{-}0.14_{-0.25}^{+0.23}$ & \hphantom{0}279.4 & 215 \\
   & on & $-0.22_{-0.14}^{+0.08}$ & $-0.15_{-0.10}^{+0.10}$ & $-0.51_{-0.12}^{+0.12}$ & $-0.80_{-0.18}^{+0.12}$ & $-0.26_{-0.19}^{+0.16}$ & \hphantom{0}261.5 & 215 \\
  \hline
  \multirow{2}{*}{$500 - 1500$} & off & $-0.58_{-0.23}^{+0.17}$ & $\hphantom{-}0.10_{-0.20}^{+0.12}$ & $-0.28_{-0.10}^{+0.14}$ & $-0.02_{-0.20}^{+0.21}$ & $\hphantom{-}0.89_{-0.30}^{+0.40}$ & \hphantom{0}214.3 & 215 \\
   & on & $-0.24_{-0.15}^{+0.29}$ & $\hphantom{-}0.14_{-0.17}^{+0.14}$ & $-0.44_{-0.12}^{+0.21}$ & $-0.20_{-0.28}^{+0.32}$ & $\hphantom{-}0.29_{-0.88}^{+0.24}$ & \hphantom{0}234.5 & 215 \\
  \hline
\end{tabular}

  \renewcommand{\arraystretch}{1.0}
\end{table*}

We repeat the analysis, but now use the DIR redshift distributions as fit model. This yields an estimate of the offset between the mean redshift of the best-fit models $\langle z_{\rm model} \rangle_i$ and the mean redshift of the true redshift distributions $\langle z_{\rm true} \rangle_i$. These offsets
\begin{equation}
  \label{eq:DIR_bias_corr}
  \Delta z_i = \langle z_{\rm model} \rangle_i - \langle z_{\rm true} \rangle_i 
  \approx \langle z_{\rm CC} \rangle_i - \langle z_{\rm true} \rangle_i
\end{equation}
(presented in Table~\ref{tab:shifts_DIR} and Fig.~\ref{fig:DIR_shifts})
are then our estimate of the shift of the mean clustering redshifts $\langle z_{\rm CC} \rangle_i$ with respect to the truth. We note that these DIR redshift distributions have been corrected for biases in the mean redshift \citep[see][]{Wright20a}. Even if the models were not corrected and significantly biased instead, this procedure would still serve as a test to detect discrepancies between the model and the clustering redshifts since they would suffer from different systematic effects (see \citeauthor{Hildebrandt20b} \citeyear{Hildebrandt20b} for a toy model).

The shift parameters obtained for the DIR and the idealised setup are within $|\Delta z| = 0.01$ for the 1st, 2nd, and 5th tomographic bin.
This is not true for the third and fourth bin where the offset is $|\Delta z_{3,4}| \gtrsim 0.02$.
We attribute this to the fact that the DIR redshift distributions are broadened compared to the true
redshift distributions due to photometric noise (compare the blue lines in Figs.~\ref{fig:true_realistic_model_plot}
and~\ref{fig:DIR_realistic_model_plot}). This broadening reduces the
skewness in these two DIR redshift
distributions, which results in a bias when these more symmetric DIR
$n(z)$ are fit to the correctly skewed clustering-$z$ data points.
The true redshift distributions in bins 1, 2, and 5 are less skewed to begin with and hence suffer less
from this DIR-specific broadening/symmetrising. The broadening also
affects the reduced $\chi^2$-values for the joint fits, which
reach up to 12, indicating a very poor fit between DIR and clustering redshift distributions.

The same poor goodness of fit can be observed when fitting the realistic mocks. There the reduced $\chi^2$-values are of
order 6, reflecting the significantly larger uncertainty at $z > 0.7$. Most shift parameter values are similar to the results from the idealised setup, but shifted by $0.01$.
This is probably driven by the strong degradation of the signal-to-noise ratio with
redshift, with the high signal-to-noise low-$z$ data points driving the fit to the broadened DIR
$n(z)$ high. The effect of the bias mitigation is comparable, both in magnitude and direction of
shift, to fitting the clustering redshifts with the true redshift distributions.

We therefore conclude that the offsets inferred with the shift-fit method are predominantly dictated by the
accuracy of (the shape of) the model redshift-distributions. In cases where the redshift distributions
are well matched to the truth, shifts are minimal. Conversely when the distribution shape is discrepant
from the truth, the shift method breaks down. The reduced $\chi^2$ may be useful as diagnostic to
detect such cases but should not be over-interpreted.

\subsection{Scale dependence and magnification} \label{sec:systematic_results}

The clustering-redshift formalism assumes linear, deterministic galaxy bias. Our fiducial scale
of \SIrange{100}{1000}{kpc} is well within the non-linear biasing regime and therefore the relation in
Eq.~(\ref{eq:crosscorr_d}) is not guaranteed to hold on such small scales. We additionally measure
clustering redshifts using the idealised reference mock sample on the slightly more conservative scales
of \SIrange{500}{1500}{kpc}. Reducing the inner radius from \SI{500}{kpc} to \SI{100}{kpc} yields an
approximate gain in signal-to-noise ratio of about \SI{50}{\percent} and we need to
test whether changes in the small scale clustering in combination with the improved sensitivity leads to
systematic shifts in our analysis. For this comparison we employ clustering redshift estimates, which are
corrected for the reference sample bias and fit with the true redshift distributions, since these exhibit the smallest statistical uncertainties

We find that the best-fit parameters (Table~\ref{tab:shifts_systematics}) are in good agreement between both scales
for the low redshift tomographic bins. For the two highest tomographic bins we detect a small positive
shift when using scales of \SIrange{500}{1500}{kpc}. We note that these shifts are still smaller than
the redshift binning of the cross-correlation measurements. The uncertainty of the $\Delta z_i$ is
roughly a factor two larger when using scales of
\SIrange{500}{1500}{kpc}, which is due to the smaller clustering
signal on large scales and the smaller logarithmic extent of the
interval \SIrange{500}{1500}{kpc} compared to
\SIrange{100}{1000}{kpc}. These larger uncertainties result in differences of
the $\Delta z_i$ between the two scales that are always insignificant at $\la 2\sigma$. In summary,
we find no strong indication that measuring on sub-\si{Mpc} scales biases the KiDS clustering redshifts
significantly. For future surveys, however, this test should be repeated, as redshift calibration
requirements will be more restrictive, and reference/target sample selections will change. 

The cross-correlation between two galaxy samples arises not only from gravitational clustering, but
also from correlations introduced by background sources that are lensed by foreground structures. This
has two effects, first it changes the effective survey area through a change in solid angle and
secondly it increases the sample depths through magnification \citep[e.g.][]{Morrison12,Choi16}. This
additional correlation may become dominant at the tails of the clustering redshifts where the
overlap between the reference and target samples is low \citep{Gatti18}. Since the shift-fitting is
mostly insensitive to changes in the tails of the distributions, we expect that magnification has
little impact on our results.

To test this assertion we repeat the clustering redshift measurements and the model fitting on a version of our
MICE2 mock catalogues in which we have switched off all lensing magnification effects. The corrections are
measured using the true galaxy positions and the galaxy samples are selected from galaxy colours and
photometric redshifts that are based on magnitudes with no flux magnification applied. The shift
parameters (Table~\ref{tab:shifts_systematics}) from both samples agree within their respective
uncertainties. The systematic shifts induced by lensing magnification in particular is not a
concern for KiDS clustering redshifts since they are significantly smaller than the KiDS
redshift calibration goal of $|\Delta z| < 0.01$.


\section{Discussion} \label{sec:discussion}

In the following we discuss the results from Sect.~\ref{sec:results} and highlight some of the challenges that need to be overcome for clustering-$z$ to become a fully complementary tool for redshift estimation.

The results we find in Sect.~\ref{sec:bias_method_results} are very encouraging. We are able to constrain the redshifts bias of the KV450 like
mock galaxies within $|\Delta z_i| \leq 0.006$ when applying the shift-fit with the true redshift
distributions. These figures may even improve with upcoming data releases which allow us to utilise
more overlap of KiDS with SDSS and 2dFLenS. However, the aforementioned figures are sensitive to
systematic features in the redshift distributions that serve as a model for the shift-fit. When we fit
the clustering redshift data points with the DIR redshift distributions, we see shifts of up to
$|\Delta z_i| \lesssim 0.04$.
These shifts are enhanced by the broadening of the DIR $n(z)$, which is induced by photometric
noise. This broadening in combination with the redshift-dependent signal-to-noise ratio of the
clustering-$z$ based on the realistic mock data yields a significant bias. In future KiDS analyses
we will utilise SOM redshift distributions \citep{Wright20a} for our shift fit analyses.
These redshift distributions are more robust than their DIR counterparts, demonstrating reduced
photometric broadening and overall bias. Using the SOM redshift distributions as a fit model, we expect, will
therefore improve the modelling of clustering redshift estimates.

We find qualitatively different behaviour between our recovered $\Delta z_i$ estimates when
calibrating with idealised and realistic mock reference samples. We hypothesise that this is
driven by the two following effects.
First, the bias evolution of the idealised reference sample is small since its amplitude
varies only by approximately \SI{25}{\percent} over the full redshift baseline. Secondly, the bias evolution has
very little impact on the redshift distribution if it is sufficiently narrow. If the bias changes by \SI{25}{\percent} over the full redshift range this also means that it only changes by $\sim \SI{5}{\percent}$ over each of the tomographic bins, an effect that is lost in the noise.
Even if there is an outlier population of galaxies e.g. at high redshift, correcting for the evolving galaxy bias may well change the true mean redshift significantly, but
the shift-fitting is mostly unaffected, since the model is not flexible enough to account for high-$z$ outliers. This explains why the idealised
mock setup is stable against changes of the sample bias. The realistic mock, however, has more
complex clustering properties since it utilises a mixture of different reference samples. Thus, the
spectroscopic bias correction has a significant impact, as can be seen by comparing the blue and orange data points in the right-hand panel of Fig.~\ref{fig:true_shifts}.

We therefore conclude that it is sufficient, at the sensitivity of KV450, to solely correct the
reference sample bias in clustering-$z$ estimates. This assessment, however, is dependent on the
clustering properties of the target sample. As a result, this conclusion will need to be revisited
in future KiDS-like analyses that utilise different source-sample selections \citep[such as those
that may be induced by additional colour-based selections;][]{Wright20a}.
The same applies to stage-IV surveys. The challenging redshift calibration requirements of these
programs will likely demand a careful treatment of the target sample bias evolution.

In the future these results can be improved upon by optimising the combination of the results from different reference samples.
The KiDS footprint and the calibration fields overlap with a rich set of spectroscopic surveys. This
allows us to calibrate redshifts to $z = 1$ and beyond, but adds the additional challenge of combining
independent measurements into an unbiased redshift estimate. We solve this problem by employing a
bootstrap resampling combination method (see Appendix~\ref{sec:pipeline}). The main issue with this approach
is that it violates a basic assumption of bootstrap resampling: the individual spatial regions
(defined by KiDS pointings and the deep spectroscopic pointings) are not statistically equivalent. Each of
the spectroscopic reference samples has a different density, redshift distribution, and clustering
properties. Thus, the number of surveys (and hence also pointings) that contribute to the combined cross-correlation amplitude
vary for the 22 redshift bins (in case of the realistic mock setup). This is problematic especially if the clustering of the reference galaxies
varies over a short redshift interval, such as in the transition region between the wide surveys (which dominate the $z \lesssim 0.7$ regime) and the deep spectroscopic surveys (which dominate the high redshifts). Figure~\ref{fig:true_realistic_model_plot}
shows a systematic feature at exactly this redshift that biases the resulting clustering
redshifts. Fortunately, this is not reflected in the shift parameters since the fit is only sensitive to
the overall shape of the redshift distributions due to the very restricted model we are fitting.

Stage-IV surveys, such as Euclid or the Vera C. Rubin Observatory Legacy Survey of Space and Time (LSST), will be in a similar situation as KiDS since they likewise
overlap with a number of (mostly, but not necessarily) spectroscopic reference samples that have distinct
properties. Therefore, additional efforts are vital in order to make clustering redshifts a
competitive and complementary method to meet the strict requirements of these projects. This can be achieved by either exploring other combination methods or by optimising the reference samples such that they bridge the gap between low-redshift, wide area surveys and high-redshift surveys with low area coverage.

Another fundamental issue for the clustering-$z$ method is correcting for the galaxy bias evolution of the target sample. The bias evolution estimates we obtain from the SBM do not agree with the results from directly
fitting the bias model to the auto-correlation terms. By design the SBM picks up any other redshift-dependent systematic error that skews the full source sample in a different way than the weighted sum of the individually normalised tomographic bin measurements. On the other hand, recovering the bias evolution requires a sufficiently accurate bias model. Furthermore, individual tomographic bins can have a slightly different bias evolution due to their selection criteria. Hence, a perfect agreement between both methods is probably too much to expect.

Regardless of the underlying cause for the disagreement, the left-hand panel of Fig.~\ref{fig:true_shifts} shows that this disagreement in the galaxy bias evolution estimates for the target sample has a vanishing impact on the recovered redshift bias parameters $\Delta z_i$. This can be explained by the fact that the bias of the KiDS mock sample evolves very little with redshift. Even a somewhat inaccurate correction does not have a strong influence on the end result.
The issue becomes more evident when applied to the realistic mock
clustering redshifts (right-hand panel of Fig.~\ref{fig:true_shifts}). Due to the low reference sample galaxy densities at high redshifts, the
signal-to-noise ratio drops significantly at $z > 0.7$. The highest tomographic redshift bin, which
contributes essential information to determine the bias evolution via the SBM, is not well constrained by the
clustering redshifts. Consequently, the uncertainty of the bias corrected redshifts and the
shift parameters $\Delta z_i$ is greatly increased, dominating the total error budget.

The only solution to this problem seems to be a more comprehensive spectroscopic calibration sample at high redshift that can reliably probe the core of the redshift distributions of the highest-redshift tomographic bins used in cosmic shear studies.


\section{Summary and outlook} \label{sec:summary}

In this paper we detail the creation of mock catalogues based on the MICE2 simulation that closely resemble the KiDS-VIKING-450 (KV450) dataset and its overlapping spectroscopic calibration samples. We use this mock data to replicate the clustering redshift estimates for KV450 and estimate their accuracy in direct comparison to the true redshifts in the simulation. The main result is that clustering-$z$ with KV450-like quality can reliably calibrate residual biases in the redshift distribution of typical galaxy samples used in cosmic shear measurements if the shape of the redshift distribution is a priori well known.
After correcting for evolving galaxy bias of a realistic spectroscopic reference sample via a measurement of its auto-correlation function, the clustering-$z$ recover the mean redshifts of all five tomographic bins at better than $|\Delta z_i| <0.006$. Without this correction, the highest-$z$ tomographic bin shows a bias of $\Delta z_5 \approx 0.026$, underlining the importance of the bias modelling of the spectroscopic sample.

Further correcting for the evolving galaxy bias of the target sample, constrained by comparing a weighted sum of the $n(z)$ of all five tomographic bins (individually normalised) to the full source sample, does not lead to a further reduction in the biases. This indicates that the very mild bias evolution of the KiDS source galaxies does not need to be corrected at this level of precision.

Next we used redshift distributions estimated from multi-colour photometry by re-weighting a deep spectroscopic calibration sample \citep[determined using the direct calibration method of the form presented by][]{Lima08}
to constrain the high redshift tails of the clustering redshifts and to interpret them as probability distributions.
Even with an idealised reference sample for the cross-correlation measurements, these noisy DIR redshift distributions are not fully able to model the clustering redshifts, due to a systematic shape mismatch between both distributions. This is exacerbated when using a more realistic reference sample, yielding biases of up to $|\Delta z_i| \approx 0.04$.
The difference, seen between our results when using the true and the DIR redshift distributions as our fit model, demonstrates an important conclusion for clustering redshift calibration.
When performing the shift-fitting with a model that is the same shape as the true redshift distribution, the resulting best-fit solution is a good representation of the truth, regardless of the model bias itself.
Therefore, these biases may be reduced by fitting redshift distributions estimated via the less noisy and less biased SOM redshift distributions of \citet{Wright20a}, or alternatively by increasing the amount of information extracted from the clustering redshifts.
A possible approach would be to use fit-models that are more flexible than a fixed redshift distribution with a single free parameter.
Such models must be constrained to positive amplitudes, as for example the Gaussian mixtures that we
employed in \citet{Hildebrandt20a}. However, fitting more complex models requires a better understanding
of the covariances of clustering redshifts, especially when derived from a combination of
different reference samples. In the long run, exploiting the full potential of synergies between
clustering- and photometry-based redshift estimates \citep[e.g.][]{Sanchez18,Alarcon20} seems to be
the most promising strategy to meet the stringent redshift requirements of upcoming stage IV survey
missions.


\begin{acknowledgements}
  We  acknowledge  support  from  the  European  Research  Council
  under grant numbers 770935 (JvdB, HH, AHW). HH is also supported by
  a Heisenberg grant (Hi1495/5-1) of the Deutsche Forschungsgemeinschaft.
  This material is based upon work supported in part by the National Science Foundation through Cooperative Agreement 1258333 managed by the Association of Universities for Research in Astronomy(AURA), and the Department of Energy under Contract No. DE-AC02-76SF00515 with the SLAC National Accelerator Laboratory. Additional LSST funding comes from private donations, grants to universities, and in-kind support from LSSTC Institutional Members. CBM acknowledges support from the DIRAC Institute in the Department of Astronomy at the University of Washington. The DIRAC Institute is supported through generous gifts from the Charles and Lisa Simonyi Fund for Arts and Sciences, and the Washington Research Foundation.
  CH acknowledges support from the European Research Council under grant number 647112, and support from the Max Planck Society and the Alexander von Humboldt Foundation in the framework of the Max Planck-Humboldt Research Award endowed by the Federal Ministry of Education and Research.
  KK acknowledges support by the Alexander von Humboldt Foundation.
  We are grateful to the zCOSMOS team to give us early access to additional deep spectroscopic redshifts that were not available in the public domain.
  This work is based on observations made with ESO Telescopes at the La Silla Paranal Observatory under programme IDs 100.A-0613, 102.A-0047, 179.A-2004, 177.A-3016, 177.A-3017, 177.A-3018, 298.A-5015, and on data products produced by the KiDS consortium.
  GAMA is a joint European-Australasian project based around a spectroscopic campaign using the Anglo-Australian Telescope. The GAMA input catalogue is based on data taken from the Sloan Digital Sky Survey and the UKIRT Infrared Deep Sky Survey. Complementary imaging of the GAMA regions is being obtained by a number of independent survey programmes including GALEX MIS, VST KiDS, VISTA VIKING, WISE, Herschel-ATLAS, GMRT and ASKAP providing UV to radio coverage. GAMA is funded by the STFC (UK), the ARC (Australia), the AAO, and the participating institutions. The GAMA website is \url{http://www.gama-survey.org/}.
  The MICE simulations have been developed at the MareNostrum supercomputer (BSC-CNS) thanks  to grants AECT-2006-2-0011 through AECT-2015-1-0013.
  This work has made use of CosmoHub \citep{Carretero17}. CosmoHub has been developed by the Port d'Informació Científica (PIC), maintained through a collaboration of the Institut de Física d'Altes Energies (IFAE) and the Centro de Investigaciones Energéticas, Medioambientales y Tecnológicas (CIEMAT), and was partially funded by the "Plan Estatal de Investigación Científica y Técnica y de Innovación" program of the Spanish government.
  \\
  {\it Author Contributions.} All authors contributed to the development and writing of this paper. The authorship list is given in three groups: the lead authors (JLvdB, HH, AHW, CBM), followed by two alphabetical groups. The first alphabetical group includes those who are key contributors to both the scientific analysis and the data products. The second group covers those who have either made a significant contribution to the data products or to the scientific analysis.
\end{acknowledgements}


\bibliographystyle{aa}
\bibliography{./references} 


\begin{appendix}

\section{Spectroscopic Mock Sample Selection} \label{sec:mock_selection}

\begin{figure}[t]
  \centering
  \includegraphics[width=0.95\columnwidth]{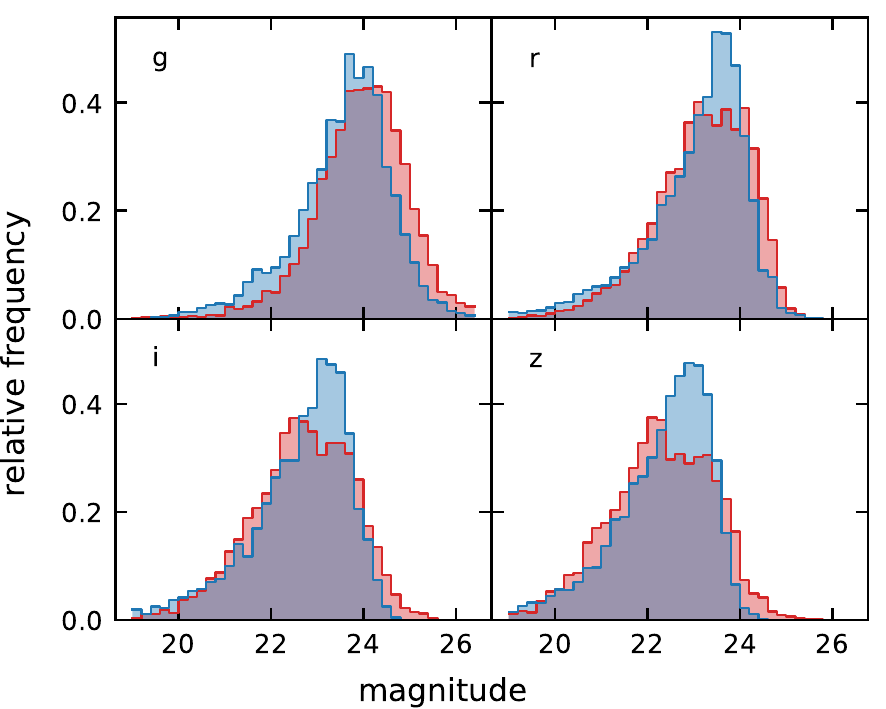}
  \\\vspace{2mm}
  \includegraphics[width=0.95\columnwidth]{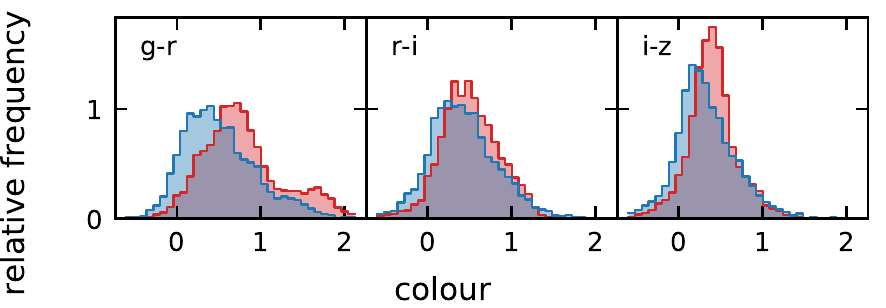}
  \\\vspace{2mm}
  \includegraphics[width=0.95\columnwidth]{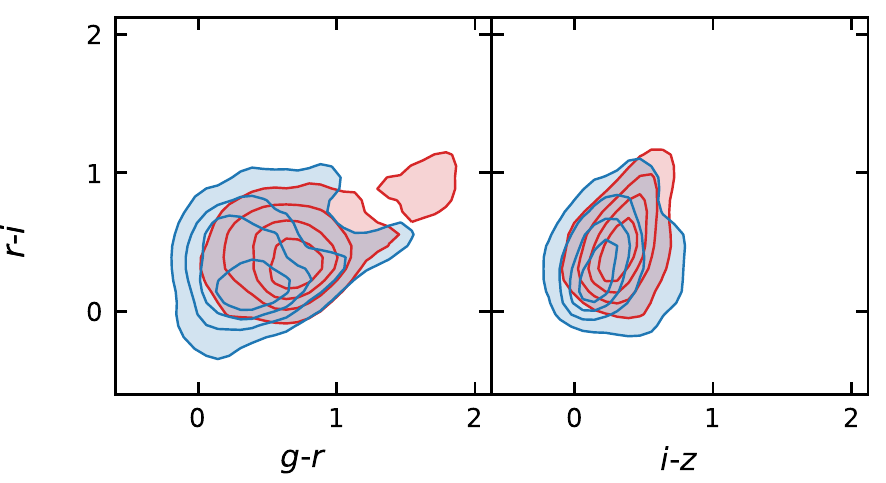}
  \caption{
    Comparison of colour and magnitude distributions for the simulated (blue) and observed (red) VVDS-02h dataset.}
  \label{fig:VVDSf02_comp}
\end{figure}

\begin{figure}[t]
  \centering
  \includegraphics[width=0.95\columnwidth]{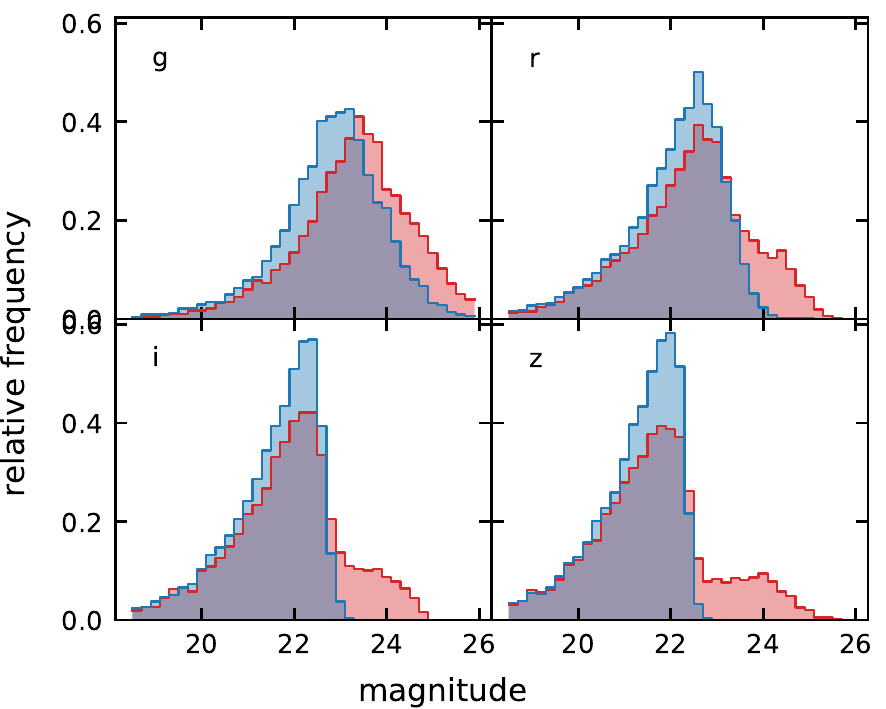}
  \\\vspace{2mm}
  \includegraphics[width=0.95\columnwidth]{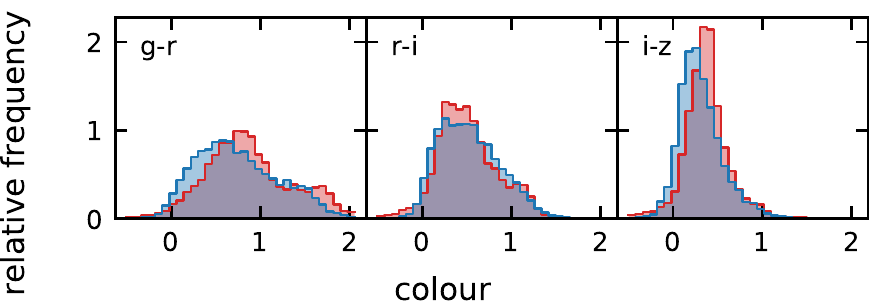}
  \\\vspace{2mm}
  \includegraphics[width=0.95\columnwidth]{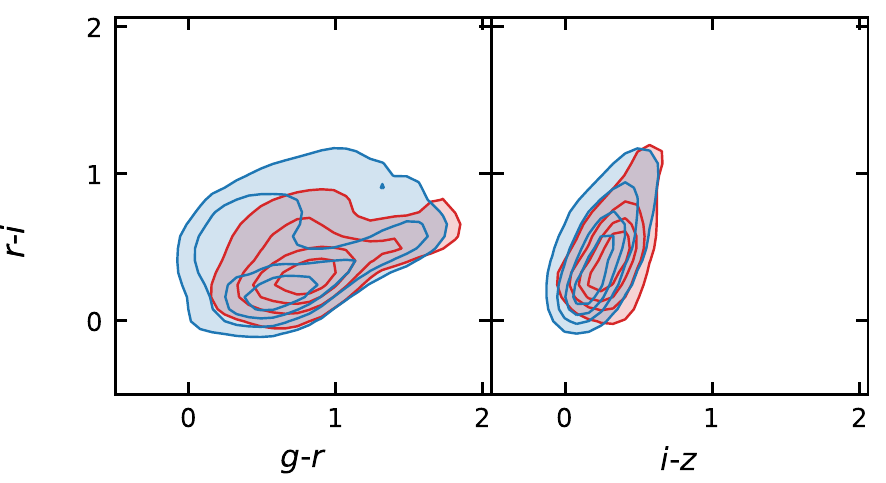}
  \caption{
    Same as Fig.~\ref{fig:VVDSf02_comp} but for the zCOSMOS dataset. The missing tails in the $i$- and $Z$-band originate
    from the zCOSMOS galaxies with $z > 1.4$ which do not exist in MICE2.}
  \label{fig:zCOSMOS_comp}
\end{figure}

Here we present the spectroscopic selection functions and their modifications for selecting these samples on MICE2 as described in Sect.~\ref{sec:mock_spec}: SDSS in presented in Table.~\ref{tab:spec_SDSS}, 2dFLenS in Table.~\ref{tab:spec_2dFLenS}, WiggleZ in Table~\ref{tab:spec_WiggleZ} and DEEP2 in Table~\ref{tab:spec_DEEP2}. 
Furthermore we show the remaining magnitude-, colour- and colour-colour plots comparing the spectroscopic data to the mock samples of VVDS-02h (Fig.~\ref{fig:VVDSf02_comp}) and zCOSMOS (Fig.~\ref{fig:zCOSMOS_comp}).

\begin{table*}
  \centering
  \caption{
    Summary of the selection functions applied in MICE2 compared to the literature selection functions, for the 
    SDSS Main galaxy sample, the BOSS CMASS and LOWZ samples, and the SDSS QSO sample.
    All selections here invoke `and' logic: ${\rm rule_1 \& rule_2 \&}$ etc. A long dash (---) indicates a selection which cannot be applied to MICE2. Deliberate adjustments that yield a better match in the simulated and real redshift distributions are highlighted in bold-face.}
  \label{tab:spec_SDSS}
  \begin{tabular}{m{0.12\textwidth}m{0.22\textwidth}m{0.22\textwidth}m{0.3\textwidth}}
    \hline\hline
    Sub-sample & SDSS Selection & MICE2 Object Selection & Comments \\
    \hline
    \multirow{1}{*}{Main}  & $r_{\rm pet} < 17.77$ & $r < 17.7$ \\
    \hline
    \multirow{3}{*}{LOWZ}  & $16.0 < r < 19.6$              & $16.0 < r < \mathbf{20.0}$               & \\
                           & $|c_\perp| < 0.2$              & $|c_\perp| < 0.2$                        & \\
                           & $r < 13.5 + c_\parallel / 0.3$ & $r < \mathbf{13.35} + c_\parallel / 0.3$ & \\
    \hline
    \multirow{4}{*}{CMASS} & $17.5 < i < 19.9$                 & $17.5 < i < \mathbf{20.1}$                          & \\
                           & $d_\perp > 0.55$                  & $d_\perp > 0.55$                                    & \\
                           & $i < 19.86 + 1.6 (d_\perp - 0.8)$ & $i < \mathbf{19.98} + 1.6 (d_\perp - \mathbf{0.7})$ & \\
                           & $r-i < 2.0$                       & $r-i < 2.0$                                         & \\
    \hline
    \multirow{3}{*}{QSO} & --- & ${\tt flag\_central}$ == 1       & \multirow{3}{0.3\textwidth}{\small Substitute selection to compensate that MICE2 does not contain quasars.} \\
                         & --- & $\log_{10}(M_{\rm halo}) > 13.3$ & \\
                         & --- & $\log_{10}(M_\star) > 11.2$      & \\
    \hline
  \end{tabular}
\end{table*}

\begin{table*}
  \centering
  \caption{
    Summary of the selection functions applied in MICE2 compared to the literature selection functions for the 2dFLenS sample.
    All selections here invoke `and' logic: ${\rm rule_1 \& rule_2 \&}$ etc.Deliberate adjustments that yield a better match in the simulated and real redshift distributions are highlighted in bold-face.}
  \label{tab:spec_2dFLenS}
  \begin{tabular}{m{0.12\textwidth}m{0.22\textwidth}m{0.22\textwidth}m{0.3\textwidth}}
    \hline\hline
    Sub-sample & 2dFLenS Selection & MICE2 Object Selection & Comments \\
    \hline
    \multirow{3}{*}{low-$z$, Cut I} & $16.0 < r < 19.2$              & $\mathbf{16.5} < r < 19.2$               & \\
                                    & $r < 13.1 + c_\parallel / 0.3$ & $r < 13.1 + c_\parallel / \mathbf{0.32}$ & \\
                                    & $|c_\perp| < 0.2$              & $|c_\perp| < 0.2$                        & \\
    \hline
    \multirow{3}{*}{low-$z$, Cut II} & $16.0 < r < 19.5$              & $\mathbf{16.5} < r < 19.5$     & \\
                                     & $|c_\perp| > 0.45 - (g-r) / 6$ & $|c_\perp| > 0.45 - (g-r) / 6$ & \\
                                     & $g-r > 1.3 + 0.25 (r-i)$       & $g-r > 1.3 + 0.25 (r-i)$       & \\
    \hline
    \multirow{3}{*}{low-$z$, Cut III} & $16.0 < r < 19.6$              & $\mathbf{16.5} < r < 19.6$               & \\
                                      & $r < 13.5 + c_\parallel / 0.3$ & $r < 13.5 + c_\parallel / \mathbf{0.32}$ & \\
                                      & $|c_\perp| < 0.2$              & $|c_\perp| < 0.2$                        & \\
    \hline
    \multirow{4}{*}{mid-$z$} & $17.5 < i < 19.9$                 & $17.5 < i < 19.9$                          & \\
                             & $r-i < 2.0$                       & $r-i < 2.0$                                & \\
                             & $d_\perp > 0.55$                  & $d_\perp > 0.55$                           & \\
                             & $i < 19.86 + 1.6 (d_\perp - 0.8)$ & $i < 19.86 + 1.6 (d_\perp - \mathbf{0.9})$ & \\
    \hline
    \multirow{5}{*}{high-$z$} & $r-W1 < 2 (r-i)$  & $r-\mathbf{K_{\rm s}} > \mathbf{1.9} (r-i)$ & \multirow{5}{0.3\textwidth}{\small Used $r-K_{\rm s}$ as substitute for missing $r-W1$ in MICE2.} \\
                              & $r-i > 0.98$      & $r-i > 0.98$                                & \\
                              & $i-Z > 0.6$       & $i-Z > 0.6$                                 & \\
                              & $19.9 < i < 21.8$ & $19.9 < i < 21.8$                           & \\
                              & $z < 19.95$       & $z < \mathbf{19.9}$                         & \\
    \hline
  \end{tabular}
\end{table*}

\begin{table*}
  \centering
  \caption{
    Summary of the selection functions applied in MICE2 compared to the literature selection functions for the WiggleZ sample.
    All selections here invoke `and' logic: ${\rm rule_1 \& rule_2 \&}$ etc.Deliberate adjustments that yield a better match in the simulated and real redshift distributions are highlighted in bold-face.
    A long dash (---) indicates a selection which cannot be applied to MICE2.
  }
  \label{tab:spec_WiggleZ}
  \begin{tabular}{m{0.12\textwidth}m{0.22\textwidth}m{0.22\textwidth}m{0.3\textwidth}}
    \hline\hline
    Selection Type & WiggleZ Selection & MICE2 Object Selection & Comments \\
    \hline
    \multirow{6}{*}{Exclusion} & $g < 22.5$        & $g < 22.5$        & \\
                               & $i < 21.5$        & $i < 21.5$        & \\
                               & $r-i < g-r - 0.1$ & $r-i < g-r - 0.1$ & \\
                               & $r-i < 0.4$       & $r-i < 0.4$       & \\
                               & $g-r > 0.6$       & $g-r > 0.6$       & \\
                               & $r-z < 0.7 (g-r)$ & $r-z < 0.7 (g-r)$ & \\
    \hline
    \multirow{6}{*}{Inclusion} & ${\rm NUV} < 22.8$                          & ---               & \multirow{6}{0.3\textwidth}{\small UV selection is mimicked by weighted sampling to match the $n(z)$'s.} \\
                               & $20.0 < r < 22.5$                           & $20.0 < r < 22.5$ & \\
                               & ${\rm FUV}-{\rm NUV} > 1$ or no ${\rm FUV}$ & ---               & \\
                               & $-0.5 < {\rm NUV} - r < 2.0$                & ---               & \\
                               & $S/N_{\rm NUV} > 3.0$                       & ---               & \\
                               & Match within \SI{2.5}{\arcsec}              & ---               & \\
    \hline
  \end{tabular}
\end{table*}

\begin{table*}
  \centering
  \caption{
    Summary of the selection functions applied in MICE2 compared to the literature selection functions for the DEEP2 sample.
    The magnitude and colour selections are linked by `and' logic, whereas the individual colour cuts invoke `or' logic. Deliberate adjustments that yield a better match in the simulated and real redshift distributions are highlighted in bold-face.}
  \label{tab:spec_DEEP2}
  \begin{tabular}{m{0.12\textwidth}m{0.22\textwidth}m{0.22\textwidth}m{0.3\textwidth}}
    \hline\hline
    Selection Type & DEEP2 Selection & MICE2 Object Selection & Comments \\
    \hline
    \multirow{1}{*}{Magnitude} & $18.5 < R < 24.0$ & $18.5 < R < 24.0$ & \\
    \hline
    \multirow{3}{*}{Colour} & $B-R < 2.45\, (R-I) - 0.2976$ & $B-R < \mathbf{2.0}\, (R-I) - \mathbf{0.4}$ & \multirow{3}{0.3\textwidth}{\small Compensate noiseless model magnitudes and template differences, see Fig.~\ref{fig:DEEP2_selection}.} \\
                            & $R-I > 1.1$                   & $R-I > 1.1$                                 & \\
                            & $B-R < 0.5$                   & $B-R < \mathbf{0.2}$                        & \\
    \hline
  \end{tabular}
\end{table*}

\section{Clustering-redshift pipeline} \label{sec:pipeline}

The \yaw{} package is an end-to-end pipeline for clustering redshift estimation. It not only allows the user to compute clustering redshifts, but also takes care of the bias mitigation. In the following we summarise our pipeline, step-by-step, from the input data to the final, bias corrected clustering redshifts.

\begin{enumerate}
  \itemsep10pt

  \item \textbf{Spatial regions:} \label{item:regions}
  We estimate the uncertainties and covariance of our clustering redshifts using bootstrap re-sampling. Therefore, we split the input data catalogues and the spectroscopic randoms into spatial regions. For the wide area spectroscopic fields the most convenient choice in KV450 was to divide the data based on the KiDS VST pointings. For the mock data used here we mimic the pointings by creating a $20 \times 22$ grid with each cell covering \SI{0.7}{deg^2}, the mean, unmasked area of a KiDS pointing.
  
  Due to their small area the deep spectroscopic fields were treated as one region each, no matter how many VST pointings were required to cover the spectroscopic footprints. As a result, there were four spatial regions originating from the deep fields, since the two DEEP2 fields (02h/23h) were considered separate entities. The mock catalogues replicate this behaviour but instead of two smaller DEEP2 mock catalogues we use one larger contiguous catalogue.
  
  \item \textbf{Random generation:}
  We generate a uniform mock-KV450 random catalogue for each spatial region independently, effectively scaling the random density based on the mean density of each pointing. We adopt the same strategy that we use for processing the KiDS data, where this local density estimation is designed to mitigate observational density variations in the data. Since we measure correlations for each spatial region independently (see next paragraph), we are not concerned that this strategy destroys large-scale correlation modes. Finally, we clone the photometric and weight distributions for the random catalogues by randomly sampling from their distributions in the mock data catalogue. Similarly, we generate spectroscopic random catalogues, cloning the spectroscopic redshift.

  \item \textbf{Cross-correlation:}
  We measure the cross-correlation between the KV450 mock data and each of the mock spectroscopic samples within an annulus of \SIrange{100}{1000}{kpc}.\footnote{We note that \citet{Hildebrandt17} used significantly smaller scales of $\SI{30}{kpc} \leq r < \SI{300}{kpc}$ to enable a clustering-$z$ measurement from the deep fields alone.} Limiting ourselves to such small scales is necessary due to the small size of the deep spectroscopic fields. The redshift resolution of our measurement is given by the binning of the spectroscopic data: for the idealised mock setup we use 45 comoving bins and the realistic mock setup 22 comoving bins between $0.01 \leq z_{\rm spec} < 1.42$. We measure the cross-correlations for the full sample and also for each tomographic bin, weighting the galaxy pairs by the KiDS {\it lensfit}-weight.

  \item \textbf{Reference sample bias:} \label{enum:pipe_spec_bias}
  We determine the reference sample bias evolution for each cross-correlation measurement by measuring the sample auto-correlation function using the same constant comoving binning and the same physical scales. With this proxy for the bias evolution we correct the cross-correlation measurements according to Eq.~(\ref{eq:spec_bias}).

  \item \textbf{Combining measurements:} \label{enum:pipe_combine}
  So far we have independent estimates for the KiDS redshift distribution for each of the spatial regions defined above that overlap with the footprints of the spectroscopic mock samples (see Fig.~\ref{fig:zspec_footprints}). The contribution of each measurement to the combined redshift distribution varies with redshift, depending on the reference sample redshift distribution. We apply a spatial bootstrapping approach to merge the cross- and auto-correlation measurements into a single redshift distribution estimate $\tilde n_{\rm p}(z)$. First, we create create a pool of pair counts $\operatorname{DD}$ and $\operatorname{DR}$ from each spatial region for each redshift bin $z_j$. Then we sum the pair counts from all $N_{\rm reg}$ regions and re-compute the correlation estimator (from Eq.~\ref{eq:estimator})
  \begin{equation}
    \bar w(z_j) = \frac{\sum_n^{N_{\rm reg}} \operatorname{DD}_n(z_j)}{\sum_n^{N_{\rm reg}} \operatorname{DR}_n(z_j)} - 1
  \end{equation}
  for the cross- and auto-correlation functions and calculate $\tilde n_{\rm p}(z_j)$ from Eq.~(\ref{eq:spec_bias}).

  This combination method violates a basic assumption of classical bootstrapping, since the sub-samples (spatial regions) are not statistically equivalent. Each of our spectroscopic samples has a different density, redshift distribution and biasing which can potentially bias the combined clustering redshifts.

  \item \textbf{Target sample bias:}
  Since we can assume that the bias evolution of the KiDS mock galaxies is the same everywhere, except for sample variance, we apply the correction after combining the measurements. We use the self-consistency bias mitigation described in Sect.~\ref{sec:bias_fit} and assume that the sample bias is approximately given by Eq.~(\ref{eq:bias_model}). We constrain the model parameter $\alpha$ using Eq.~(\ref{eq:fit_residual}) by comparing the redshift distribution estimate of step~\ref{enum:pipe_combine} for the full KiDS mock sample and the weighted sum of the tomographic bins. The weight of a bin is given by the sum of the {\it lensfit}-weights of all galaxies in that bin, divided by the sum of the weights of all galaxies between $0.1 < z_{\rm phot} \leq 1.2$.

  We also explore one alternative way of correcting the target sample bias. Analogous to step \ref{enum:pipe_spec_bias} we compute the sample auto-correlation function using the true redshifts of the mock galaxies and combine the measurements with the bootstrap method from step \ref{enum:pipe_combine}. According to Eq.~(\ref{eq:crosscorr_d}), this should give the most accurate clustering redshifts and we use it to validate the bias fitting approach. Certainly, such an approach is not possible on real data.

  \item \textbf{Covariance Estimation:}
  To estimate uncertainties and a covariance matrix for the clustering redshifts we apply bootstrap resampling based on the spatial regions. We implement this in the same way as the survey combination in step \ref{enum:pipe_combine}, but instead of summing all regions together, we randomly draw with replacement from the pool of spatial regions to generate samples. We propagate the bias mitigation to these samples which allows us to compute uncertainties from the standard error and a covariance matrix.

\end{enumerate}

There are two key differences between \yaw{} and \thewizz{} which we used previously in \citet{Hildebrandt20a}. \yaw{} is no longer built around the pixelation library \stomp{}. Furthermore, we removed one of the most distinct features of \thewizz{}, which is its ability to create look-up-tables with galaxy indices for each pairs. This allows to compute clustering redshifts for arbitrary sub-samples of the photometric galaxies at any later time by querying the look-up table before summing the pair counts and computing the correlation estimator. Creating the look-up table is very time-consuming and our datasets are not big enough to harvest the full potential of this approach. Instead, \yaw{} stores just the pair counts for each reference source, giving the user the freedom to change the redshift binning of the correlation measurements at a later stage.

\end{appendix}


\end{document}